\documentclass[journal,11pt,left = 1.25in, right = 1.25in, top = 1in, bottom = 1in, onecolumn,draftclsnofoot,]{IEEEtran}
\usepackage{amsmath,amsfonts}
\usepackage{algorithmic}
\usepackage{algorithm}
\usepackage{array}
\usepackage{booktabs}
\usepackage[caption=false,font=normalsize,labelfont=sf,textfont=sf]{subfig}
\usepackage{textcomp}
\usepackage{stfloats}
\usepackage{url}
\usepackage{hyperref}
\usepackage{verbatim}
\usepackage{graphicx}
\usepackage{epstopdf}
\usepackage{cite}
\usepackage{todonotes}
\usepackage{caption}
\usepackage{multirow}
\usepackage{ulem}
\usepackage{tikz}
\usepackage{orcidlink}
\usetikzlibrary{calc,positioning,fit,shapes,arrows,backgrounds,intersections,math,decorations.text, patterns, angles, quotes, external, shapes.misc}
\usetikzlibrary{%
	decorations.pathreplacing,%
	decorations.pathmorphing%
}
\usetikzlibrary{arrows.meta}
\usepackage{pgfplots}
\usepackage{xcolor}
\usepackage{etoolbox}
\usepackage[flushleft]{threeparttable}
\newcommand{\orcid}[1]{\href{https://orcid.org/#1}{\includegraphics[width=8pt]{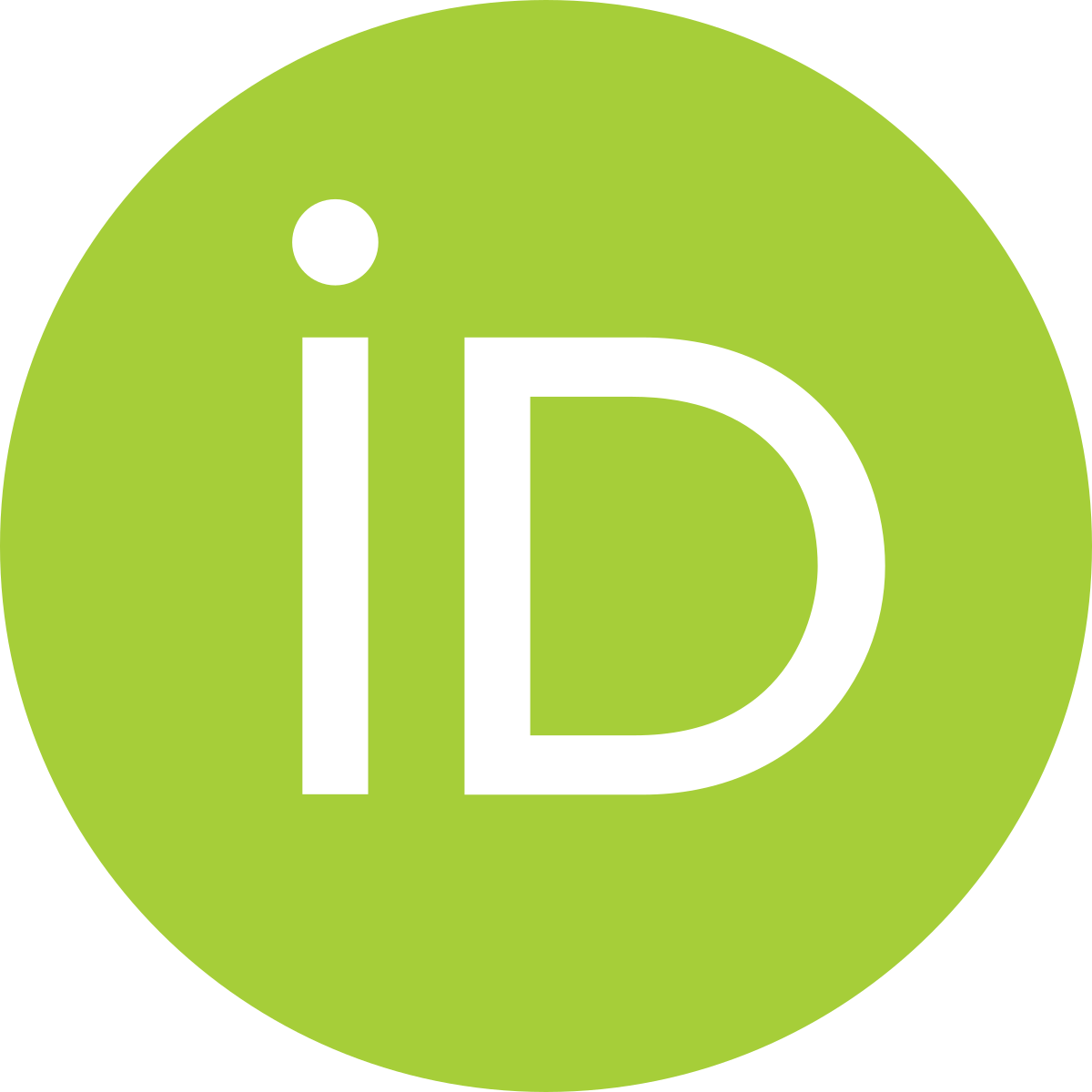}}}
\newcommand{\tj}{\text{\sf j}}
\newcommand{\tT}{\text{\sf T}}
\newcommand{\tH}{\text{\sf H}}
\newcommand{\tF}{\text{\sf F}}

\newcommand{\tst}{\text{\small{subject to}}}

\newcommand{\diag}{{\rm diag}}

\renewcommand{\vec}{\ensuremath{{\rm vec}}}

\newcommand{\tr}{{\rm Tr}}

\newcommand{\mb}[1]{\ensuremath{\mathit{\boldsymbol{#1}}}}

\newcommand{\rvb}[1]{\ensuremath{\mathit{\boldsymbol{#1}}}}
\newcommand{\bmtx}{\ensuremath{\begin{bmatrix}}}
\newcommand{\emtx}{\ensuremath{\end{bmatrix}}}

\newcommand{\sparse}[1]{\ensuremath{\tilde{#1}}}

\pgfplotsset{every axis legend/.append style={row sep=-3pt, inner ysep=0pt, inner xsep=2pt}}

\tikzset{every picture/.append style={baseline=(current bounding box.north west)}}

\colorlet{cs0}{black!30!yellow}
\colorlet{cs1}{black!30!red}
\colorlet{cs2}{white!30!blue}
\colorlet{cs3}{black!25!green}
\colorlet{cs4}{black!25!yellow}

\tikzstyle{sensor}=[anchor=center,circle, fill=cs0, inner sep=0pt, minimum height=1.5mm]
\tikzstyle{sensorLarge1}=[anchor=center, regular polygon, regular polygon sides = 3, fill=cs0, inner sep=0pt, minimum height=2.5mm]
\tikzstyle{sensorLarge2}=[anchor=center, regular polygon, regular polygon sides = 4, fill=cs0, inner sep=0pt, minimum height=2.5mm]
\tikzset{cross/.style={cross out, draw=black, minimum size=2*(#1-\pgflinewidth), inner sep=0pt, outer sep=0pt},
	cross/.default={1 mm}}
\tikzstyle{source}=[anchor=center,fill=red, inner sep=0pt, minimum height=1.5mm, minimum width=1.5mm]

\tikzstyle{sensorBg}=[line width=3mm, line join=round, cap=round, fill=black!10, draw=black!10]
\tikzstyle{mybox} = [draw=blue!60, thick, rectangle, rounded corners, inner sep=10pt, inner ysep=15pt]
\tikzstyle{myboxsmall} = [draw=blue!60, thick, rectangle, rounded corners, inner sep=10pt, inner ysep=7.5pt]
\tikzstyle{fancytitle} =[fill=blue!60, text=white, rounded corners]

\pgfdeclarelayer{background}
\pgfdeclarelayer{foreground}
\pgfsetlayers{background,main,foreground}

\def\NSnp{4}

\def\seedA{107}

\def\bSizeMod{1.7mm}

\def\seedA{20}

\newcommand{\drawMtx}[4]{
	\begin{tikzpicture}
		#1
		\pgfmathsetseed{\seedA}
		\foreach \x in {1,...,#3}{
		\foreach \y in {1,...,#2}{
			\pgfmathrandomitem{\c}{color};
			\pgfmathrandominteger{\r}{75}{100};
			\fill[\c!\r] (\x*#4, \y*#4) rectangle +(#4,#4);
		}
		}
	\end{tikzpicture}
}

\newcommand{\drawMtxB}[4]{
	\begin{tikzpicture}
		\foreach \x in {1,...,#3}{
		\foreach \y in {1,...,#2}{			
			\pgfmathrandominteger{\r}{50}{100};
			\fill[#1!\r] (\x*#4, \y*#4) rectangle +(#4,#4);
		}
		}
	\end{tikzpicture}
}

\renewcommand{\vec}[1]{\boldsymbol{#1}}
\newcommand{\EV}[1]{\mathbb{E}\left\lbrace #1\right\rbrace}

\newcommand{\realset}{\mathbb{R}}
\newcommand{\complexset}{\mathbb{C}}
\newcommand{\norm}[1]{\left|\left|#1\right|\right|}

\newcommand{\herm}{^\tH}
\newcommand{\trans}{^\tT}

\newcommand{\oproj}[2]{\vec{\Pi}_{#1}^{\perp}#2}

\newcommand{\iter}[2]{#1^{(#2)}}

\definecolor{TUDa-2b}{HTML}{0083CC}
\definecolor{TUDa-2c}{HTML}{00689D}
\definecolor{TUDa-2d}{HTML}{004E73}
\definecolor{TUDa-4a}{HTML}{AFCC50}
\definecolor{TUDa-4b}{HTML}{99C000}
\definecolor{TUDa-4c}{HTML}{7FAB16}
\definecolor{TUDa-4d}{HTML}{6A8B22}
\definecolor{TUDa-8a}{HTML}{EE7A34}
\definecolor{TUDa-8b}{HTML}{EC6500}
\definecolor{TUDa-9b}{HTML}{E6001A}
\definecolor{TUDa-11b}{HTML}{721085}
\definecolor{TUDa-11c}{HTML}{611C73}
\definecolor{TUDa-0d}{HTML}{535353}
\definecolor{TUDa-0c}{HTML}{898989}
\definecolor{TUDa-0b}{HTML}{B5B5B5}
\definecolor{TUDa-0a}{HTML}{DCDCDC}

\newdimen\XCoord
\newdimen\YCoord

\newboolean{emphasize}
\setboolean{emphasize}{false}

\newboolean{isCanvasDrawn}
\setboolean{isCanvasDrawn}{false}

\def\opacityBorder{0}
\def\borderSWX{-10.7}
\def\borderSWY{-3.5}
\def\borderNEX{3.0}
\def\borderNEY{3.0}

\definecolor{col1}{RGB}{13,83,144}
\definecolor{col2}{RGB}{8,101,168}
\definecolor{col3}{RGB}{0,133,210}
\definecolor{col4}{RGB}{0,145,222}
\definecolor{col5}{RGB}{0,164,240}
\definecolor{col6}{RGB}{12,183,236}
\definecolor{col7}{RGB}{71,205,147}
\definecolor{col8}{RGB}{121,223,72}
\definecolor{col9}{RGB}{184,179,48}
\definecolor{col10}{RGB}{255,102,51}
\definecolor{col11}{RGB}{214,78,98}
\definecolor{col12}{RGB}{176,57,141}
\definecolor{col13}{RGB}{140,36,182}
\definecolor{col14}{RGB}{107,17,220}
\definecolor{col15}{RGB}{78,1,253}
\definecolor{col16}{RGB}{57,0,196}
\definecolor{col17}{RGB}{40,0,144}
\definecolor{col18}{RGB}{30,0,115}
\definecolor{col19}{RGB}{28,0,110}

\definecolor{trueCol5}{RGB}{0,164,240}
\definecolor{trueCol8}{RGB}{121,223,72}
\definecolor{trueCol10}{RGB}{255,102,51}
\definecolor{trueCol14}{RGB}{107,17,220}

\tikzmath{
	\angSrc1 = 180 -40;
	\angSrc2 = 180 -70;
	\angSrc3 = 180 -90;
	\angSrc4 = 180 -130;
}
\def\angA{180-40}
\def\angB{180-70}
\def\angC{180-90}
\def\angD{180-130}
\def\angE{180-85}

\def\ySrc{2.4}

\def\posDictMatX{-9.45}
\def\posDictMatY{1.75}

\def\sqrSize{0.25}

\pgfdeclareverticalshading{verticalRainbow}{50bp}{%
	color(0bp)=(TUDa-11c!50!black);
	color(6bp)=(TUDa-11b);
	color(25bp)=(TUDa-8a);
	color(32bp)=(TUDa-4a!90!black);
	color(42bp)=(TUDa-2b);
	color(49bp)=(TUDa-2b!50!TUDa-2d);
	color(50bp)=(TUDa-2d!75!darkgray)
}
\pgfdeclarehorizontalshading{horizontalRainbow}{50bp}{%
	color(0bp)=(TUDa-2d!75!darkgray);
	color(3bp)=(TUDa-2b);
	color(18bp)=(TUDa-4a!90!black);		
	color(25bp)=(TUDa-8a);	
	color(44bp)=(TUDa-11b);	
	color(50bp)=(TUDa-11c!50!black)
}
\pgfdeclarehorizontalshading{horizontalLinear}{50bp}{%
	color(0bp)=(TUDa-2d!75!darkgray);
	color(10bp)=(TUDa-2b);
	color(22bp)=(TUDa-4a!90!black);		
	color(25bp)=(TUDa-8a);	
	color(40bp)=(TUDa-11b);	
	color(50bp)=(TUDa-11c!50!black)
}

\def\arcThickness{0.05}

\def\offsetSensorPos{0.5}

\def\gridSliderSize{0.75}
\tikzstyle{gridSlider} = [regular polygon, regular polygon sides = 3, draw, inner sep=\gridSliderSize pt, fill = TUDa-9b]

\def\highlightBorder{0}

\def\posReceiveSigX{-9.85}
\def\posReceiveSigY{-0.75}

\def\posTrueSteerMatX{-5.7}
\def\posTrueSteerMatY{-0.75}

\def\posTestSigX{-7.4}
\def\posTestSigY{-0.75}

\def\posSrcSigMatX{-4.45}
\def\posSrcSigMatY{-0.75}

\def\posSpecX{-2.5}
\def\posSpecY{-3.1}

\def\posEqSymX{-6.75}
\def\posEqSymY{-1.5}

\def\posDotSymX{-4.91}
\def\posDotSymY{-1}

\usepackage{hyperref}
\newcommand\copyrighttext{%
	\footnotesize \textcopyright 2023 IEEE. 
	Personal use of this material is permitted. Permission from IEEE must be obtained for all other uses, in any current or future media, including reprinting/republishing this material for advertising or promotional purposes, creating new collective works, for resale or redistribution to servers or lists, or reuse of any copyrighted component of this work in other works. 
	DOI: \href{LINK TO YOUR ARTICLE}{10.1109/MSP.2023.3255558}
}
\makeatletter
\def\ps@IEEEtitlepagestyle{
	\def\@oddfoot{\mycopyrightnotice}
	\def\@evenfoot{}
}
\def\mycopyrightnotice{
	{\footnotesize
		\begin{minipage}{\textwidth-2\fboxsep}%
			\centering%
			\noindent\fbox{\parbox{\linewidth}{\copyrighttext}}
		\end{minipage}
	}
}

\begin{document}

\title{Three more Decades in Array Signal Processing Research: An Optimization and Structure Exploitation Perspective}

\author{Marius~Pesavento,
		Minh~Trinh-Hoang
		and~Mats~Viberg
\thanks{Marius Pesavento \orcid{0000-0003-3395-2588} is with the Communication
	Systems Group, TU Darmstadt, Darmstadt, Germany
	(e-mail: {pesavento}@nt.tu-darmstadt.de).}
\thanks{Minh Trinh-Hoang \orcid{0000-0002-9162-6644} is with Rohde\&Schwarz, Munich, Germany.}
\thanks{Mats Viberg \orcid{0000-0003-1549-419X} is with Blekinge Institute of Technology, Sweden (e-mail: mats.viberg@bth.se).}
}


\maketitle

%

\section{Introduction}
\IEEEPARstart{T}{he} signal processing community currently witnesses the emergence of sensor array processing and Direction-of-Arrival (DoA) estimation in various modern applications, such as automotive radar, mobile user and millimeter wave indoor localization, drone surveillance, as well as in new paradigms, such as joint sensing and communication in future wireless systems. This trend is further enhanced by technology leaps and availability of powerful and affordable multi-antenna hardware platforms. New multi-antenna technology has led to a widespread use of such systems in contemporary sensing and communication systems as well as a continuous evolution towards larger multi-antenna systems in various application domains, such as massive Multiple-Input-Multiple-Output (MIMO) communications systems comprising hundreds of antenna elements. The massive increase of the antenna array dimension leads to unprecedented resolution capabilities which opens new opportunities and challenges for signal processing. For example, in large MIMO systems, modern array processing methods can be used to estimate and track the physical path parameters such as DoA, Direction-of-Departure (DoD), Time-Delay-of-Arrival (TDoA) and Doppler shift of tens or hundreds of multipath components with extremely high precision \cite{Gao19}. This parametric approach for massive MIMO channel estimation and characterization benefits from enhanced resolution capabilities of large array systems and efficient array processing techniques. Direction-based MIMO channel estimation, which has not been possible in small MIMO systems due to limited number of antennas, not only significantly reduces the complexity but also improves the quality of MIMO channel prediction as the physical channel parameters generally evolve on a much smaller time-scale than the MIMO channel coefficients. 

The history of advances in super resolution DoA estimation techniques is long, starting from the early parametric multi-source methods such as the computationally expensive maximum likelihood (ML) techniques to the early subspace-based techniques such as Pisarenko and MUSIC \cite{VanTrees02}. Inspired by the seminal review paper “Two Decades of Array Signal Processing Research: The Parametric Approach” by Krim and Viberg published in the IEEE Signal Processing Magazine \cite{KrimV96}, we are looking back at another three decades in Array Signal Processing Research under the classical narrowband array processing model based on second order statistics. We revisit major trends in the field and retell the story of array signal processing from a modern optimization and structure exploitation perspective. In our overview, through prominent examples, we illustrate how different DoA estimation methods can be cast as optimization problems with side constraints originating from prior knowledge regarding the structure of the measurement system. Due to space limitations, our review of the DoA estimation research in the past three decades is by no means complete. For didactic reasons, we mainly focus on developments in the field that easily relate the traditional multi-source estimation criteria in \cite{KrimV96} and choose simple illustrative examples.

As many optimization problems in sensor array processing are notoriously difficult to solve exactly due to their nonlinearity and multimodality, a common approach is to apply problem relaxation and approximation techniques in the development of computationally efficient and close-to-optimal DoA estimation methods. The DoA estimation approaches developed in the last thirty years differ in the prior information and model assumptions that are maintained and relaxed during the approximation and relaxation procedure in the optimization.

Along the line of constrained optimization, problem relaxation and approximation, recently, the Partial Relaxation (PR) technique has been proposed as a new optimization-based DoA estimation framework that applies modern relaxation techniques to traditional multi-source estimation criteria to achieve new estimators with excellent estimation performance at affordable computational complexity. In many senses, it can be observed that the estimators designed under the PR framework admit new insights in existing methods of this well-established field of research \cite{Trinh-Hoang18}.

The introduction of sparse optimization techniques for DoA estimation and source localization in the late noughties marks another methodological leap in the field \cite{Fuchs_1998,Fuchs_2004,Ender10,Stoica11,chellappa_chapter_2018}. These modern optimization-based methods became extremely popular due to their advantages in practically important scenarios where classical subspace-based techniques for DoA estimation often experience a performance breakdown, e.g., in the case of correlated sources, when the number of snapshots is low, or when the model order is unknown. Sparse representation-based methods have been successfully extended to incorporate and exploit various forms of structure, e.g., application-dependent row- and rank-sparse structures \cite{Malioutov05,Steffens18} that induce joint sparsity to enhance estimation performance in the case of multiple snapshots. In particular array geometries, additional structure, such as Vandermonde and shift-invariance, can be used to obtain efficient parameterizations of the array sensing matrix that avoid the usual requirement of sparse reconstruction methods to sample the angular Field-of-View (FoV) on a fine DoA grid \cite{Recht13,Zai16}. Despite the success of sparsity-based methods, it is, however, often neglected that these methods also have their limitations such as estimation biases resulting from off-grid errors and the impact of the sparse regularization, high computational complexity and memory demands as well as sensitivity to the choice of the so-called hyperparameters.
In fact, for many practical estimation scenarios, sparse optimization techniques are often outperformed by classical subspace techniques in terms of both resolution of sources and computational complexity. From the theoretical perspective, performance guarantees of sparse methods are generally only available under the condition of minimum angular separation between the source signals \cite{chellappa_chapter_2018}. 
Therefore, it is important to be aware of these limitations and to appreciate the benefits of both traditional and modern optimization-based DoA estimation methods.

The narrowband far-field point source signals with perfectly calibrated sensor arrays and centralized processing architectures have been fundamental assumptions in the past. With the trend of wider reception bandwidth on the one hand, and larger aperture and distributed array on the other hand, the aforementioned assumptions appeared restrictive and often impractical. Distributed sensor networks have emerged as a scalable solution for source localization where sensors exchange data locally within their neighborhood and in-network processing is used for distributed source localization with low communication overhead \cite{Scaglione08}. Furthermore, DoA estimation methods for partly calibrated subarray systems have been explored \cite{Pesavento02,See04}.

Model structure, e.g., in the form of favorable spatial sampling pattern, is exploited for various purposes: either to reduce the computational complexity and to make the estimation computationally tractable, or to generally improve the estimation quality. In this paper, we revisit the major trends of structure exploitation in sensor array signal processing. Along this line, we consider advanced spatial sampling concepts designed in recent years, including minimum redundancy \cite{Moffet68}, augmentable \cite{Abramovich98}, nested \cite{Pal10} and co-prime arrays \cite{Tan14,Qin15}. The aforementioned spatial sampling patterns were designed to facilitate new DoA estimation methods with the capability of resolving significantly more sources than sensors in the array. This is different from conventional sampling pattern, e.g., ULA, where the number of identifiable sources is always smaller than the number of sensors.

\section{Signal model}\label{subsect:signal_model}
In this overview paper, we consider the narrowband point source signal model\footnote{In practical wireless communication or radar applications, the receive signal may be broadband. Such scenarios require extensions of the narrowband signal model, e.g., to subband processing or the multidimensional harmonic retrieval, which is however out of scope of this paper.}. Under this signal model, we are interested in estimating the DoAs, i.e., the parameter vector $\mb{\theta} = [\theta_1,\ldots,\theta_N]^\tT$, of $N$ far-field narrow-band sources impinging on a sensor array comprised of $M$ sensors from noisy measurements. We assume that the DoA $\theta_n$ lies in the FoV $\Theta$, i.e., $\theta_n \in\Theta$. Let $\mb{x}(t) = \mb{A}(\mb{\theta})\mb{s}(t)+\mb{n}(t)$ denote the linear array measurement model at time instant $t$ where $\mb{s}(t)$ and $\mb{n}(t)$ denote the signal waveform vector and the sensor noise vector, respectively. The sensor noise $\mb{n}(t)$ is commonly assumed to be a zero-mean spatially white complex circular Gaussian random process with covariance matrix $\nu \mb{I}_M$.
The steering matrix $\mb{A}(\mb{\theta}) \in \mathcal{A}_N$ lives on a $N$-dimensional array manifold $\mathcal{A}_N$, which is defined as
\begin{equation}
	\label{eq:defManifold}
\mathcal{A}_N = \left\lbrace\ \mb{A} = \left[\mb{a}(\vartheta _1),\ldots,  \mb{a}(\vartheta _N)\right] \big| \vartheta_1 < \ldots< \vartheta_N \text{ and } \vartheta_n\in\Theta\text{ for all } n= 1, \ldots, N\right\rbrace.
\end{equation}In \eqref{eq:defManifold}, the steering vector $\mb{a}(\theta) = [e^{-\tj\pi d_1 \cos(\theta)},$ $e^{-\tj\pi d_2 \cos(\theta)},\ldots, e^{-\tj\pi d_M \cos(\theta)} ]^\tT$ denotes, e.g., the array response of a linear array with sensor positions $d_1, \ldots, d_M$ in half-wavelength for a narrow-band signal impinging from the direction $\theta$. The steering matrix $\mb{A}(\mb{\theta}) = \left[\mb{a}(\theta_1), \ldots, \mb{a}(\theta_N)\right]$ must satisfy certain regularity conditions so that the estimated DoAs can be uniquely identifiable up to a permutation from the noiseless measurement. Mathematically, the unique identifiability condition requires that if $\mb{A}( \mb{\theta}^{(1)} ) \mb{s}^{(1)}(t) = \mb{A}( \mb{\theta}^{(2)} ) \mb{s}^{(2)}(t)$ for $t = 1, \ldots, T$ then $\mb{\theta}^{(1)}$ is a permutation of $\mb{\theta}^{(2)}$. Generally, this condition must be verified for any sensor structure and the corresponding FoV. Specifically, it can be shown that if the array manifold is free from ambiguities, i.e., if any oversampled steering matrix $\mb{A}(\mb{\theta}) \in \mathcal{A}_K$ of dimension $M\times K$ with $K \geq M$ has a Kruskal rank $q\big( \mb{A}(\mb{\theta}) \big) = M$, then  $N$ DoAs with $N < M$ can be uniquely determined from the noiseless measurement \cite{KrimV96}. Equivalently, any set of $M$ column vectors $\left\lbrace\mb{a}(\theta_1), \ldots, \mb{a}(\theta_M)\right\rbrace$ with $M$ distinct DoAs $\theta_1, \ldots, \theta_M \in \Theta$ are linearly independent. 
In the so-called conditional signal model, the waveform vector $\mb{s}(t)$ is assumed to be deterministic such that $\rvb{x}(t) \sim {\cal N}_{\rm C}\big( \mb{A}(\mb{\theta}) \mb{s}(t), \nu \mb{I}_M\big)$. The unknown noise variance $\nu$ and the signal waveform $\mb{S} = [\mb{s}(1),\ldots, \mb{s}(T)]$ are generally not of interest in the context of DoA estimation, but they are necessary components of the signal model. In contrast, in the unconditional signal model, the waveform is assumed to be zero-mean complex circular Gaussian  such that $\rvb{x}(t) \sim {\cal N}_{\rm C}\big( \mb{0}_{M}, \mb{A}(\mb{\theta}) \mb{P} \mb{A}^\tH(\mb{\theta}) + \nu \mb{I}_M\big)$, where the noise variance $\nu$ and the waveform covariance matrix $\mb{P}= \EV{\mb{s}(t)\mb{s}^\tH(t)}$ are considered as unknown parameters. We assume, if not stated otherwise, that the signals are not fully correlated, i.e., $\mb{P}$ is nonsingular. 
\section{Cost function and concentration}
\begin{figure}
 	\centering
%
%
\begin{tikzpicture}
\begin{axis}[%
width=5.5cm,
height=5.5cm,
at={(0.877in,0.481in)},
scale only axis,
point meta min=0,
point meta max=31.6491081028761,
axis on top,
xmin=-0.497237569060773,
xmax=179.502762430939,
ymin=-0.497237569060773,
ymax=179.502762430939,
every tick label/.append style={font=\scriptsize},
title= {\small $M = 10, \boldsymbol{\theta} = \left[105^\circ, 120^\circ\right]\trans, \text{SNR} = 0\text{ dB}, T = 100$},
title style = {at={(0.5, 0.97)}}, 
xtick = {45,9 0, 135},
xticklabels = {$45^\circ$, $90^\circ$, $135^\circ$}, 
ytick = {45, 90, 135},
yticklabels = {$45^\circ$, $90^\circ$, $135^\circ$}, 
xlabel={\small $\theta_1$},
x label style = {at={(axis description cs: 0.5, 0.05)}},
ylabel={\small $\theta_2$},
y label style = {at={(axis description cs: 0.08, 0.5)}},
axis background/.style={fill=white},
legend style={legend cell align=left, align=left, draw=white!15!black, row sep=-2pt, at={(0.02,0.98)}, anchor = north west, font=\fontsize{8}{5}\selectfont},
colormap={mymap}{[1pt] rgb(0pt)=(0.2422,0.1504,0.6603); rgb(78pt)=(0.237836,0.306245,0.830674); rgb(131pt)=(0.178747,0.475359,0.884032); rgb(163pt)=(0.0843976,0.623041,0.817465); rgb(188pt)=(0.0341301,0.736629,0.750443); rgb(189pt)=(0.0473088,0.740162,0.740694); rgb(190pt)=(0.0625298,0.743579,0.730846); rgb(191pt)=(0.078935,0.746892,0.720903); rgb(192pt)=(0.0956101,0.750119,0.71087); rgb(193pt)=(0.111783,0.753284,0.700748); rgb(194pt)=(0.12695,0.756413,0.690531); rgb(195pt)=(0.14085,0.759528,0.680206); rgb(196pt)=(0.153396,0.762647,0.669749); rgb(197pt)=(0.164637,0.765779,0.659127); rgb(198pt)=(0.174741,0.768925,0.648297); rgb(199pt)=(0.183984,0.772077,0.637216); rgb(200pt)=(0.192743,0.775215,0.625849); rgb(201pt)=(0.201448,0.778311,0.614171); rgb(202pt)=(0.210519,0.781329,0.602164); rgb(203pt)=(0.220321,0.784233,0.58982); rgb(204pt)=(0.231125,0.78699,0.577147); rgb(205pt)=(0.243118,0.78957,0.564154); rgb(206pt)=(0.256415,0.791945,0.550833); rgb(207pt)=(0.271155,0.794086,0.537097); rgb(208pt)=(0.287484,0.795991,0.522706); rgb(209pt)=(0.305801,0.797669,0.507135); rgb(210pt)=(0.3268,0.799119,0.489603); rgb(211pt)=(0.351246,0.800288,0.469352); rgb(212pt)=(0.379288,0.801051,0.4463); rgb(213pt)=(0.410214,0.801245,0.42123); rgb(214pt)=(0.443101,0.80073,0.395103); rgb(215pt)=(0.477153,0.799438,0.368618); rgb(216pt)=(0.51159,0.79738,0.342276); rgb(217pt)=(0.548406,0.794299,0.314439); rgb(218pt)=(0.587037,0.79018,0.285658); rgb(219pt)=(0.626053,0.785124,0.257665); rgb(220pt)=(0.665797,0.779047,0.231524); rgb(226pt)=(0.847325,0.797812,0.171178); rgb(228pt)=(0.895281,0.814868,0.161702); rgb(229pt)=(0.914957,0.826722,0.158089); rgb(236pt)=(0.967513,0.917252,0.123248); rgb(237pt)=(0.969704,0.928362,0.117547); rgb(238pt)=(0.971219,0.938969,0.111809); rgb(240pt)=(0.973005,0.955973,0.1017); rgb(243pt)=(0.974529,0.970779,0.0914714); rgb(244pt)=(0.974974,0.974537,0.0886997); rgb(246pt)=(0.975725,0.979428,0.0847974); rgb(247pt)=(0.975981,0.980691,0.0837007); rgb(252pt)=(0.976539,0.98275,0.081708); rgb(255pt)=(0.97676,0.983473,0.0809688)},
colorbar
]
\addplot [forget plot] graphics [xmin=-0.497237569060773, xmax=179.502762430939, ymin=-0.497237569060773, ymax=179.502762430939] {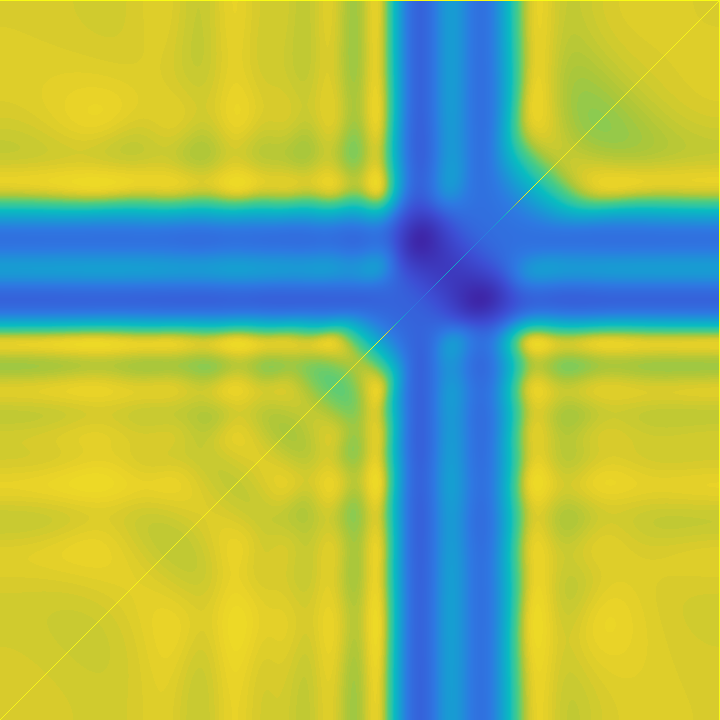};
\addplot [color=red, draw=none, mark=o, mark options={solid, red, line width = 0.75pt}, mark size = 2.5pt, only marks]
  table[row sep=crcr]{%
142.25	0\\
76.75	6.25\\
88.25	6.25\\
50.75	7\\
65.5	7\\
104.75	16.25\\
22	21.75\\
21.75	22\\
179.5	27.25\\
120.5	28.5\\
141.25	31.5\\
65.75	33.25\\
88.25	33.5\\
76.75	34\\
43	42.75\\
42.75	43\\
179.5	50.5\\
7	50.75\\
120.5	50.75\\
141	50.75\\
88.25	53.25\\
76.75	54.5\\
104.75	56.25\\
60	59.75\\
59.75	60\\
7	65.5\\
141	65.5\\
179.5	65.5\\
33.25	65.75\\
120.5	66\\
88.25	66.5\\
104.75	67.75\\
72.25	72\\
72	72.25\\
141.25	76.5\\
6.25	76.75\\
34	76.75\\
54.5	76.75\\
179.5	76.75\\
84.25	84\\
84	84.25\\
104.75	85.5\\
120.5	87.5\\
141.5	88\\
6.25	88.25\\
33.5	88.25\\
53.25	88.25\\
66.5	88.25\\
179.5	88.25\\
16.25	104.75\\
56.25	104.75\\
67.75	104.75\\
85.5	104.75\\
142	104.75\\
179.5	104.75\\
120	105\\
105	120\\
28.5	120.5\\
50.75	120.5\\
66	120.5\\
87.5	120.5\\
137.25	120.5\\
179.5	120.5\\
124.5	124.25\\
124.25	124.5\\
120.5	137.25\\
50.75	141\\
65.5	141\\
31.5	141.25\\
76.5	141.25\\
88	141.5\\
104.75	142\\
0	142.25\\
149.75	149.5\\
149.5	149.75\\
179.5	179.25\\
27.25	179.5\\
50.5	179.5\\
65.5	179.5\\
76.75	179.5\\
88.25	179.5\\
104.75	179.5\\
120.5	179.5\\
179.25	179.5\\
};
\addlegendentry{\scriptsize Local Minima}

\addplot [color=green, draw=none, mark=x, mark options={solid, green, line width = 1pt}, only marks, mark size = 3pt]
  table[row sep=crcr]{%
105	120\\
120	105\\
};
\addlegendentry{\scriptsize True DoAs}
\end{axis}
\node[anchor = north, align = center] at (9.4cm, 1.2cm){\scriptsize DML\\[-11pt]\scriptsize Cost Function};
\end{tikzpicture}%
 	\caption{Example of the DML cost function for two sources evaluated over the FoV. Multiple local minima are observed. Consequently, local optimization search cannot guarantee to converge to the global minimum.}
 	\label{figure:MultimodalObjective}
 \end{figure}
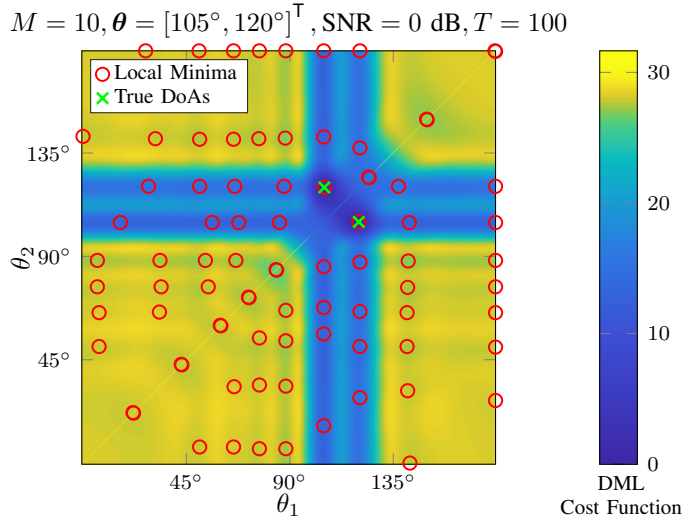

Parametric methods for DoA estimation can generally be cast as optimization problems with multivariate objective functions that depend on a particular data matrix $\mb{Y}$ obtained from the array measurements $\mb{X} = [\mb{x}(1),\ldots,\mb{x}(T)]$ through a suitable mapping, the unknown DoA parameters of interest $\mb{\theta}$ and the unknown nuisance parameters, which we denote by the vector $\mb{\alpha}$. Hence, the parameter estimates are computed as the minimizer of the corresponding optimization problem with the objective function $f(\mb{Y}|\mb{A}(\mb{\theta}),\mb{\alpha})$ as follows
\begin{equation}
\label{eq:generalDOAProblem}
\mb{A}(\hat{\mb{\theta}}) = \underset{\mb{A}(\mb{\theta})\in\mathcal{A}_N}{\arg\min}
\underset{\mb{\alpha}}{\min}~f(\mb{Y}| \mb{A}(\mb{\theta}), \mb{\alpha}).
\end{equation}
Remark that in \eqref{eq:generalDOAProblem}, we make no restriction how the data matrix $\mb{Y}$ is constructed from the measurement matrix $\mb{X}$. For example, in the most trivial case, the data matrix $\mb{Y}$ can directly represent the array measurement matrix, i.e., $\mb{Y} = \mb{X}$. However, for other optimization criteria, the data matrix $\mb{Y}$ can be the sample covariance matrix, i.e., $\mb{Y} = \hat{\mb{R}} = \dfrac{1}{T}\mb{X}\mb{X}^\tH$ as a sufficient statistics, or even the signal eigenvectors $\mb{Y} = \hat{\mb{U}}_\text{s}$ (or the noise eigenvectors $\mb{Y} = \hat{\mb{U}}_\text{n}$) obtained from the eigendecomposition $\hat{\mb{R}} = \hat{\mb{U}}_\text{s} \hat{\mb{\Lambda}}_\text{s} \hat{\mb{U}}_\text{s}^\tH + \hat{\mb{U}}_\text{n} \hat{\mb{\Lambda}}_\text{n} \hat{\mb{U}}_\text{n}^\tH$ where $\hat{\mb{\Lambda}}_\text{s} = \diag \big(\hat{\lambda}_1, \ldots, \hat{\lambda}_N \big)$ contains the $N$-largest eigenvalues of $\hat{\mb{R}}$.
In Table \ref{tab:Estimators}, some prominent examples of multi-source estimation methods are listed: Deterministic Maximum Likelihood (DML)\cite[Sec. 8.5.2]{VanTrees02}, Weighted Subspace Fitting (WSF)\cite{Viberg91} and COvariance Matching Estimation Techniques (COMET)\cite{Ottersten1998}. \\
As we are primarily interested in estimating the DoA parameters $\mb{\theta}$, a common approach is to concentrate the objective function with respect to all (or only part of) the nuisance parameters $\mb{\alpha}$. In case that a closed-form minimizer of the nuisance parameters w.r.t the remaining parameters exists, the expression of this minimizer can be inserted back to the original objective function to obtain the concentrated optimization problem. More specifically, let $\hat{\mb{\alpha}}(\mb{\theta})$ denote the minimizer of the full problem for nuisance parameter vector $\mb{\alpha}$ as a function of $\mb{\theta}$, i.e., $\hat{\mb{\alpha}}(\mb{\theta}) = \underset{\mb{\alpha}}{\arg\min}~f(\mb{Y}| \mb{A}(\mb{\theta}), \mb{\alpha})$. The concentrated objective function $g\big(\mb{Y}|\mb{A}(\mb{\theta}) \big) = f\big(\mb{Y}|\mb{A}(\mb{\theta}),\hat{\mb{\alpha}}( \mb{\theta} )\big)$ then only depends on the DoAs $\mb{\theta}$. Apart from the reduction of dimensionality, the concentrated versions of multi-source optimization problems often admit appealing interpretations. In Table \ref{tab:Estimators}, the concentrated criteria corresponding to the previously considered full-parameter multi-source criteria are provided. We observe, e.g., in the case of the concentrated DML and the WSF criteria that at the optimum, the residual signal energy contained in the nullspace of the steering matrix is minimized. 
\\
Due to the complicated structure of the array manifold $\mathcal{A}_N$ in \eqref{eq:defManifold}, the concentrated objective function $g\big(\mb{Y}|\mb{A}(\mb{\theta}) \big)$ is, for common choices in Table \ref{tab:Estimators}, highly non-convex and multi-modal w.r.t. the DoA parameters $\mb{\theta}$. Consequently, the concentrated cost function contains a large number of local minima in the vicinity of the global minimum. This can, e.g., be observed in Figure~\ref{figure:MultimodalObjective} where the cost function of the DML estimator is depicted. While multi-source estimation criteria generally show unprecedented asymptotic as well as threshold performance for low sample size, Signal-to-Noise Ratio (SNR) and closely spaced sources, their associated computational cost is unsuitable in many practical applications. The exact minimization generally requires an $N$-dimensional search over the FoV which becomes computationally prohibitive even for low source numbers, e.g., $N = 3$. 
 
In the past three decades and beyond, significant efforts have been made to devise advanced DoA estimation algorithms that exhibit good trade-offs between performance and complexity. While some very efficient methods have been proposed in a different context and based on pure heuristics, in this feature article, we focus on optimization-based estimators that stem, in some way or the other, from multi-source optimization problems for the classical array processing model, c.f. Table \ref{tab:Estimators}. Considering the array processing literature, a vast amount of estimators proposed in past years can be derived from multi-source optimization problems. 
Optimization-based estimators have the advantage that they are not only well-motivated but also intuitively interpretable and flexible for generalization to more sophisticated realistic array signal models. Interestingly, some of the estimators in Table \ref{tab:Estimators} were initially derived by heuristics and are reintroduced here from the perspective of multi-source optimization problems.
\section{Modern convex optimization for DoA estimation}
The progress in modern convex optimization theory and the emergence of efficient constrained optimization solvers with the turn of the millennium, such as, e.g., the SeDuMi software for solving semi-definite programs, had significant impact on the research across disciplines in the signal processing, communication and control communities. In fact, it comes at no surprise that the advances in sensor array signal processing of the past three decades are well aligned with this trend that facilitates advanced constrained optimization-based design approaches.  
Three closely related universal concepts have been intensively used in array signal processing to make optimization-based estimation procedures numerically stable and computationally feasible. These are: \emph{i)} structure exploitation, \emph{ii)} approximation, and \emph{iii)} relaxation.

\emph{Structure exploitation} refers to techniques that make use of particular redundancies in the measurement system to introduce convenient data reorganizations and reparameterizations. Examples are methods particularly designed for uniform, shift-invariant and co-prime array geometries.
 
\emph{Problem approximation} techniques provide local approximations of the multidimensional multi-modal nonconvex objective function with the goal to decompose a complex problem into several subproblems. Each subproblem, whose minimizer is much generally simpler to obtain than that of the original problem, is solved in parallel or sequentially, ideally in closed-form. Examples are the Expectation-Maximization algorithm, the orthogonal matching pursuit, and the single-source approximation methods.

\emph{Problem relaxation} techniques in DoA estimation aim at simplifying the complicated manifold structure associated with the estimation problem. The manifold relaxation is carried out, e.g., to convexify the constraint sets in the associated optimization problems such that numerical methods can be applied. 

Approximation and relaxation techniques have in common that they are used to deliberately ignore some parts of the problem structure at the expense of optimality or performance of the solution. The objective is to simplify the problem so that efficient suboptimal solutions can be obtained that in many cases are close to optimal and often even admit performance guarantees. The DoA estimators reviewed in this overview paper apply one or more of the aforementioned optimization concepts, as explained in more detail in the following subsections.
\subsection{Single Source Approximation}
\label{subsec:SingleSrcApprox}
\begin{figure}[h]
	\centering
	\input{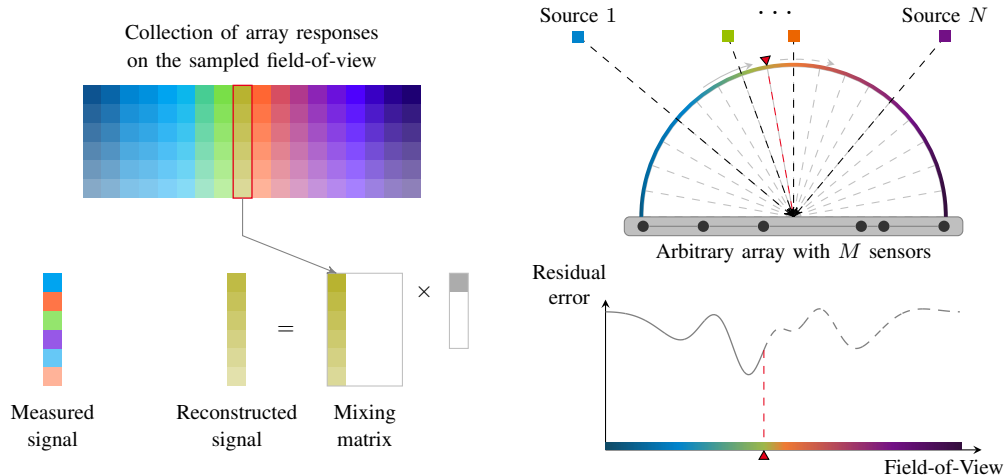}
	\caption{Illustration of the  Single Source Approximation concept: The optimization is carried out only over a single-source model with array manifold, i.e., $\mb{A}(\mb{\theta}) \rightarrow \mb{a}(\theta) \in \mathcal{A}_1$ (note that the mixing matrix has only one non-zero column corresponding to the candidate DoA $\mb{a}(\theta)$). The influence of the remaining source signals during the spectral search is neglected, which is denoted by zero columns in the mixing matrix. The data term $\mb{Y}$ in the objective function of the single source approximation method is however identical to that of the corresponding multi-source optimization problem.}
	\label{fig:SingleSrcIllustration}
\end{figure}

Spectral-based DoA estimation methods like the popular MUSIC algorithm belong to the class of single source approximation methods. In contrast to the full parameter search of minimizing the multi-source objective $f\big(\mb{Y}|\mb{A}(\mb{\theta}),\mb{\alpha}\big)$ over the $N$-source signal model with array manifold $\mathcal{A}_N$ and nuisance parameter vector $\mb{\alpha}$, the optimization problem in the single source approximation approach is simplified and the optimization is carried out only over a single source model with array manifold, i.e., $\mb{A}(\mb{\theta}) \rightarrow \mb{a}(\theta) \in \mathcal{A}_1$ and nuisance parameters $\mb{\alpha} \rightarrow \mb{\alpha}_1$. It is important to note that, while the number of signal components considered in the optimization is reduced in the single source approximation approach, the data term $\mb{Y}$ in the objective remains unchanged. The locations $\mb{a}(\hat{\theta})$ of the $N$-deepest minima of the so-called null-spectrum $f\big( \mb{Y}|\mb{a}(\theta),\mb{\alpha}_1 \big)$ evaluated for all steering vectors $\mb{a}(\theta) \in \mathcal{A}_1$ with angles in the FoV, are considered as the steering vector of the estimated DoAs. By using the compact notation $^N\arg\min~g(\cdot)$ to represent the spectral search of the cost function $g(\cdot)$ for the $N$-deepest local minima, the single source approximation is formulated as follows:
\begin{align}
	\big\{ \mb{a}(\hat{\theta}) \big\}  & = 	\underset{\mb{a}(\theta)\in\mathcal{A}_1}{^N\arg\min}\hspace*{2pt}\underset{\mb{\alpha}_1}{\min} \ f\big( \mb{Y}|\mb{a}(\theta),\mb{\alpha}_1 \big).
\end{align}
For clarity, the concept of the single source approximation and the corresponding spectral search is visualized in Figure~\ref{fig:SingleSrcIllustration}. As summarized in Table~\ref{tab:Estimators}, classical spectral search methods such as the Conventional Beamformer, Capon beamformer and MUSIC can be reformulated as single source approximations of the corresponding multi-source criteria.

\subsection{Partial Relaxation Methods} \label{section:PR_methods}
\begin{figure}
	\centering
	\input{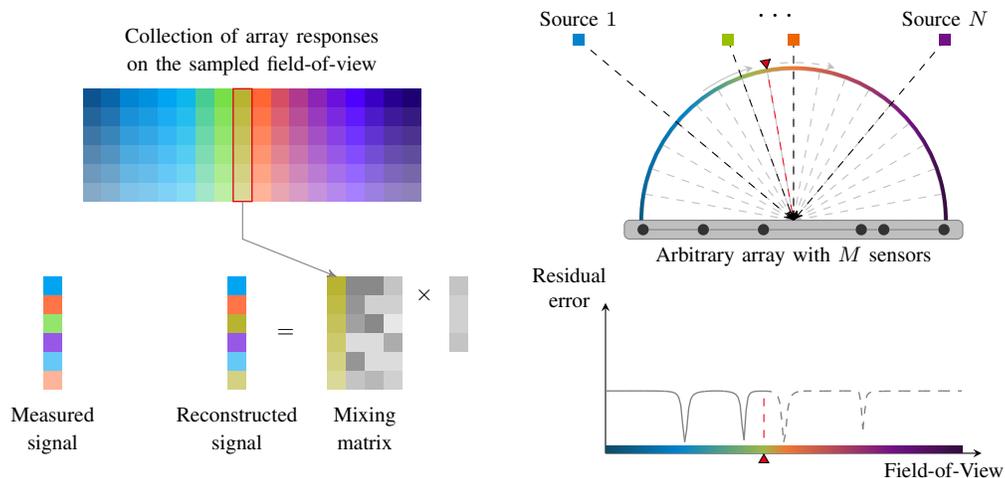}
	\caption{Illustration of the  Partial Relaxation concept: The optimization is carried out over the relaxed array manifold $\bar{\mathcal{A}}_N$, where the structure of the first column in $\mb{a}(\theta_1)$ is maintained and the structure of the remaining columns is relaxed to an arbitrary complex matrix, $\left[\vec{a}(\theta_2), \ldots, \vec{a}(\theta_N)\right] \rightarrow \mb{B}\in\complexset^{M\times (N-1)}$. Unlike the single source approximation, the influence of the remaining source signals during the spectral search is considered by the unstructured matrix $\mb{B}$ (depicted by gray columns in the mixing matrix), which generally leads to an improvement of the DoA estimation when sources are closely spaced.}
	\label{fig:PRIllustration}
\end{figure}

Similar to the conventional parametric methods, the Partial Relaxation (PR) approach considers the signals from all potential source directions in the multi-source cost function. However, to make the problem tractable, the array structures of some signal components are relaxed. More precisely, instead of enforcing the steering matrix ${\vec{A} = \left[\vec{a}(\theta_1), \ldots, \vec{a}(\theta_N)\right]}$ to be an element in the highly structured array manifold $\mathcal{A}_N$ as in the multi-source criteria in \eqref{eq:generalDOAProblem}, without the loss of generality, we maintain the manifold structure of only the first column $\vec{a}(\theta_1)$ of $\vec{A}$, which corresponds to the signal of consideration. On the other hand, the manifold structure of the remaining sources $\left[\vec{a}(\theta_2), \ldots, \vec{a}(\theta_N)\right]$, which are considered as interfering sources, is relaxed to an arbitrary matrix ${\vec{B}\in\complexset^{M\times(N-1)}}$ \cite{Trinh-Hoang18}. Mathematically, we assume that $\vec{A}\in\bar{\mathcal{A}}_N$ where the relaxed array manifold $\bar{\mathcal{A}}_N$ is parameterized as:
\begin{equation}
\bar{\mathcal{A}}_N = \left\lbrace \vec{A}(\vartheta) = \left[\vec{a}(\vartheta), \vec{B}\right]\big| \vec{a}(\vartheta)\in\mathcal{A}_1, \vec{B}\in\complexset^{M\times(N-1)}\right\rbrace.
\label{eq: relaxedArrayManifold}
\end{equation}
We remark that every matrix element in the relaxed array manifold $\bar{\mathcal{A}}_N$ in \eqref{eq: relaxedArrayManifold} still retains the specific structure from the geometry of the sensor array in its first column, hence the name partial relaxation. However, only one DoA can be estimated from the first column of the matrix minimizer if the cost function of \eqref{eq:generalDOAProblem} is minimized on the relaxed array manifold $\bar{\mathcal{A}}_N$ of \eqref{eq: relaxedArrayManifold}. Therefore, we perform the spectral search similarly to the single source approximation in Subsection~\ref{subsec:SingleSrcApprox} as follows: first we fix the data matrix $\vec{Y}$, minimize and concentrate the objective function in \eqref{eq:generalDOAProblem} with respect to $\vec{B}$ and other nuisance parameters $\mb{\alpha}$ to obtain the concentrated cost function. Then, we evaluate the concentrated cost function for different values of $\vec{a}(\vartheta)\in\mathcal{A}_1$ to determine the locations of the $N$-deepest local minima. The concept of the PR approach is illustrated in Figure~\ref{fig:PRIllustration}. Using similar notation as in the single source approximation approach, the PR approach admits the following general optimization problem
\begin{equation}
\begin{aligned}
\big\{ \mb{a}(\hat{\theta}) \big\} &= \underset{\mb{A}(\theta)\in\bar{\mathcal{A}}_N}{^N\arg\min}~f\big( \mb{Y}|\mb{A}(\theta),\mb{\alpha} \big).\\
&= 	\underset{\mb{a}(\theta)\in\mathcal{A}_1}{^N\arg\min}~\underset{\mb{B}\in\complexset^{M\times (N-1)}}{\min}~\underset{\mb{\alpha}}{\min} ~f\big( \mb{Y}|\left[\mb{a}(\theta), \mb{B}\right],\mb{\alpha} \big).
\end{aligned}
\label{eq:PRGeneralForm}
\end{equation}
The rationale for the PR approach lies in the fact that, when a candidate DoA $\theta$ coincides with one of the true DoAs $\theta_n$, then with $\vec{B}$ modeling the steering vectors of the remaining DoAs, a perfect fit to the data is attained at high SNR or large number of snapshots $T$. When the candidate $\theta$ is however different from all true DoAs $\theta_n$, the number of degrees-of-freedom in $\vec{B}$ is not sufficiently large to represent the contribution of all $N$ source signals. By applying different cost functions to the general optimization problem in \eqref{eq:PRGeneralForm}, multiple novel estimators in the  PR framework are obtained in \cite{Trinh-Hoang18}. A summary of estimators under the PR framework and their relation with conventional multi-source and single source approximation-based DoA estimators is provided in Table~\ref{tab:Estimators}.

\begin{table}
	\def\widthFirstCol{1.25}
	\def\widthSecondCol{4.25}
	\def\widthThirdCol{5.5}
	\def\widthFourthCol{4.25}
	\def\widthMainLine{1}
	\def\widthSubLine{1}
	\begin{threeparttable}
		\caption{Conventional DoA Estimators}\label{tab:Estimators}
		\begin{tabular}{ccccc}
			\noalign{\hrule height \widthMainLine pt}
			\multicolumn{2}{c}{\small \textbf{}}                                                          & \small\textbf{Full Dimension} & \small\textbf{Partial Relaxation} & \small\textbf{Single Source Approx.} \\ \noalign{\hrule height \widthMainLine pt}
			
			\multicolumn{1}{l}{\multirow{2}{*}{\scriptsize \begin{tabular}[c]{@{}l@{}} \\\rotatebox[origin=c]{90}{\textbf{Signal Fitting}}\end{tabular}}} & \scriptsize \rotatebox[origin=c]{90}{Original} & \parbox{\widthSecondCol cm}{\scriptsize \begin{equation*}
				\underset{\substack{\vec{A}\in\mathcal{A}_N\\ \vec{S}\in\complexset^{N\times T}}}{\arg\min}\norm{\vec{X}- \vec{A}\vec{S}}_{\tF}^2
				\end{equation*}}  & \parbox{\widthThirdCol cm}{\scriptsize \begin{equation*}
				\underset{\vec{a}\in\mathcal{A}_1}{^N\arg\min}\text{ }\underset{\vec{s},\vec{B}, \vec{J}}{\min} \text{ } \norm{\vec{X} - \vec{a}\vec{s}\trans - \vec{B}\vec{J}}_{\tF}^2
				\end{equation*}}  &  \parbox{\widthFourthCol cm}{\scriptsize \begin{equation*}
				\underset{\vec{a}\in\mathcal{A}_1}{^N\arg\min}~\underset{\vec{s}\in\complexset^{T}}{\min}\norm{\vec{X}- \vec{a}\vec{s}\trans}_{\tF}^2
				\end{equation*}} \\ \cline{2-5} 
			\multicolumn{1}{l}{}                                                               & \scriptsize \rotatebox[origin=c]{90}{Concentr.} & \parbox{\widthSecondCol cm}{\scriptsize \begin{equation*}
				\underset{\vec{A}\in\mathcal{A}_N}{\arg\min}~\text{tr}\left(\oproj{\vec{A}}{\vec{X}\vec{X}\herm}\right)
				\end{equation*}}  & \parbox{\widthThirdCol cm}{\scriptsize \begin{equation*}
				\underset{\vec{a}\in\mathcal{A}_1}{^N \arg\min}~\sum_{k = N}^M \lambda_k\left(\oproj{\vec{a}}{\vec{X}\vec{X}\herm}\right)
				\end{equation*}}  & \parbox{\widthFourthCol cm}{\scriptsize \begin{equation*}
				\underset{\vec{a}\in\mathcal{A}_1}{^N \arg\min}~\text{tr}\left(\oproj{\vec{a}}{\vec{X}\vec{X}\herm}\right)
				\end{equation*}}  \\ \multicolumn{1}{l}{}                                                               & \scriptsize  & \parbox{\widthSecondCol cm}{\centering \scriptsize DML}  & \parbox{\widthThirdCol cm}{\centering \scriptsize PR-DML}  & \parbox{\widthFourthCol cm}{\centering Conventional Beamformer}\\\noalign{\hrule height \widthSubLine pt}
			
			\multicolumn{1}{l}{\multirow{2}{*}{\scriptsize \begin{tabular}[c]{@{}l@{}} \\\rotatebox[origin = c]{90}{\textbf{Subspace Fitting}}\end{tabular}}} & \scriptsize \rotatebox[origin=c]{90}{Original} & \parbox{\widthSecondCol cm}{\scriptsize \begin{equation*}\hspace*{-3pt}
				\underset{\substack{\vec{A}\in\mathcal{A}_N,\\ \vec{V}\in\complexset^{N\times N}}}{\arg\min}\norm{\hat{\vec{U}}_\text{s}\vec{W}^{\frac{1}{2}}- \vec{A}\vec{V}}_{\tF}^2
				\end{equation*}}  & \parbox{\widthThirdCol cm}{\scriptsize \begin{equation*}
				\underset{\vec{a}\in\mathcal{A}_1}{^N\arg\min}~\underset{\textit{\textbf{v}},\vec{B}, \vec{Q}}{\min}{}\norm{\hat{\vec{U}}_\text{s}\vec{W}^{\frac{1}{2}} - \vec{a}\textbf{\textit{v}}\trans - \vec{B}\vec{Q}}_{\tF}^2
				\end{equation*}}  &  \parbox{\widthFourthCol cm}{\scriptsize \begin{equation*} \underset{\vec{a}\in\mathcal{A}_1}{^N\arg\min}~\underset{\textit{\textbf{v}}\in\complexset^{N}}{\min}\norm{\hat{\vec{U}}_\text{s}\vec{W}^{\frac{1}{2}}- \vec{a}\textbf{\textit{v}}\trans}_{\tF}^2
				\end{equation*}} \\ \cline{2-5} 
			\multicolumn{1}{l}{}                                                               & \scriptsize \rotatebox[origin=c]{90}{Concentr.} & \parbox{\widthSecondCol cm}{\scriptsize \begin{equation*}
				\underset{\vec{A}\in\mathcal{A}_N}{\arg\min}~\text{tr}\left(\oproj{\vec{A}}{\hat{\vec{U}}_\text{s}\vec{W}\hat{\vec{U}}\herm_\text{s}}\right)
				\end{equation*}}  & \parbox{\widthThirdCol cm}{\scriptsize \begin{equation*}
				\underset{\vec{a}\in\mathcal{A}_1}{^N \arg\min}~\sum_{k = N}^M \lambda_k\left(\oproj{\vec{a}}{\hat{\vec{U}}_\text{s}\vec{W}\hat{\vec{U}}_\text{s}\herm}\right)
				\end{equation*}}  & \parbox{\widthFourthCol cm}{\scriptsize \begin{equation*}
				\underset{\vec{a}\in\mathcal{A}_1}{^N \arg\min}~\text{tr}\left(\oproj{\vec{a}}{\hat{\vec{U}}_\text{s}\vec{W}\hat{\vec{U}}_\text{s}\herm}\right)
				\end{equation*}}  \\  
			\multicolumn{1}{l}{}                                                               & \scriptsize  & \parbox{\widthSecondCol cm}{\centering \scriptsize WSF}  & \parbox{\widthThirdCol cm}{\centering \scriptsize PR-WSF}  & \parbox{\widthFourthCol cm}{\centering Variant of Weighted MUSIC\tnote{*}}\\\noalign{\hrule height \widthSubLine pt}

			\multicolumn{1}{l}{\multirow{2}{*}{\scriptsize \begin{tabular}[c]{@{}l@{}} \\[-5pt]\rotatebox[origin = c]{90}{\textbf{Covariance Fitting}}\end{tabular}}} & \scriptsize \rotatebox[origin=c]{90}{Original} & \parbox{\widthSecondCol cm}{\scriptsize \begin{gather*}\hspace*{-3pt}
				\begin{aligned}
				\underset{\substack{\vec{A}\in\mathcal{A}_N\\ } }{\arg\min}&\norm{\hat{\vec{R}}^{-\tfrac{1}{2}}\left(\hat{\vec{R}} - \vec{R}\right)\hat{\vec{R}}^{-\tfrac{1}{2}}}_{\tF}^2\\
				\text{s.t. } & \vec{R} = \vec{A}\vec{P}\vec{A}\herm + \nu\vec{I}
				\end{aligned}
				\end{gather*}}  & \parbox{\widthThirdCol cm}{\scriptsize \begin{gather*}
				\begin{aligned}
				\underset{\vec{a}\in\mathcal{A}_1}{^N\arg\min}\underset{\sigma_\text{s}^2\geq 0,\vec{G}}{\min} &\norm{\hat{\vec{R}} - \sigma_\text{s}^2\vec{a}\vec{a}\herm - \vec{G}\vec{G}\herm}_{\tF}^2\\[-5pt]
				\text{s.t. } &\hat{\vec{R}} - \sigma_s^2\vec{a}\vec{a}\herm - \vec{G}\vec{G}\herm \succeq \vec{0}\\[-5pt]
				&\text{rank}(\vec{G})\leq N-1
				\end{aligned}\raisetag{0.75\baselineskip}
				\end{gather*}}  &  \parbox{\widthFourthCol cm}{\scriptsize \begin{gather*}
				\begin{aligned}
				\underset{\vec{a}\in\mathcal{A}_1}{^N\arg\min}\text{ }\underset{\sigma_s^2\geq 0}{\min}&\norm{\hat{\vec{R}} - \sigma_s^2\vec{a}\vec{a}\herm}_{\tF}^2\\[-5pt]		
				\text{s.t. } &\hat{\vec{R}} - \sigma_s^2\vec{a}\vec{a}\herm\succeq \vec{0}
				\end{aligned}
				\end{gather*}} \\ \cline{2-5} 
			\multicolumn{1}{l}{}                                                               & \scriptsize \rotatebox[origin=c]{90}{Concentr.} & \parbox{\widthSecondCol cm}{\scriptsize \centering See \cite[Eq. (35)]{Ottersten1998}}  & \parbox{\widthThirdCol cm}{\scriptsize \begin{equation*}
				\underset{\vec{a}\in\mathcal{A}_1}{^N \arg\min}~\sum_{k = N}^M \lambda_k^2\left(\hat{\vec{R}} - \dfrac{1}{\vec{a}\herm\hat{\vec{R}}^{-1}\vec{a}}\vec{a}\vec{a}\herm\right)
				\end{equation*}}  & \parbox{\widthFourthCol cm}{\scriptsize \begin{equation*}
				\underset{\vec{a}\in\mathcal{A}_1}{^N \arg\min}~\norm{\hat{\vec{R}} - \dfrac{1}{\vec{a}\herm\hat{\vec{R}}^{-1}\vec{a}}\vec{a}\vec{a}\herm}_{\tF}^2
				\end{equation*}}  \\ \multicolumn{1}{l}{}                                                               & \scriptsize  & \parbox{\widthSecondCol cm}{\centering \scriptsize COMET}  & \parbox{\widthThirdCol cm}{\centering \scriptsize PR-CCF}  & \parbox{\widthFourthCol cm}{\centering Variant of Capon Beamformer\tnote{**}}\\\noalign{\hrule height \widthSubLine pt}
			
		\end{tabular}
		\begin{tablenotes}
			\item $\boldsymbol{\Pi}_{\boldsymbol{A}}^\perp = {\boldsymbol I} - \boldsymbol{A} (\boldsymbol{A}^{\text{\sf{H}}} \boldsymbol{A})^{-1}\boldsymbol{A}^{\text{\sf{H}}}$ denotes the orthogonal projector onto the nullspace spanned by the columns of $\boldsymbol{A}$. The code for the different variants of the PR methods can be downloaded at \url{https://github.com/PartialRelaxationMethods}.
			\item[*] Conventional weighted MUSIC algorithm (e.g., see \cite[Eq. (9.258)]{VanTrees02}) applies the weighting on the noise subspace. This variant applies the weighting on the signal subspace.
			\item[**] Note that the optimizer $\hat{\sigma}_\text{s}^2$ is the spectrum of the Capon Beamformer. The null-spectrum of this estimator contains both the spectra of the Conventional Beamformer and the Capon Beamformer
		\end{tablenotes}
	\end{threeparttable}
	
\end{table}

\subsection{Sequential Techniques}\label{subsection:iterative_methods}
\begin{figure}
	\centering	
%
%
\definecolor{mycolor1}{RGB}{136, 0, 21}%
\definecolor{mycolor2}{RGB}{185,122,187}%
\definecolor{mycolor3}{RGB}{255,32,27}%
\definecolor{mycolor4}{RGB}{255,174,201}%
\definecolor{mycolor5}{RGB}{255,137,29}%
\definecolor{mycolor6}{RGB}{255,201,14}%
\definecolor{mycolor7}{RGB}{255,242,0}%
\definecolor{mycolor8}{RGB}{239,228,176}%
\definecolor{mycolor9}{RGB}{31,194,43}%
\definecolor{mycolor10}{RGB}{181,230,29}%
\definecolor{mycolor11}{RGB}{0,119,251}%
\definecolor{mycolor12}{RGB}{94,243,255}%

\definecolor{mycolor13}{RGB}{63,72,204}%
\definecolor{mycolor14}{RGB}{112,146,190}%
\definecolor{mycolor15}{RGB}{163,73,164}%
\definecolor{mycolor16}{RGB}{200,191,231}%
\begin{tikzpicture}

\begin{axis}[%
height=0.36\linewidth, 
width=0.54\linewidth, 
at={(0in,0in)},
scale only axis,
every tick label/.append style={font=\scriptsize},
xmode=log,
xmin=10,
xmax=10000,
xlabel={\small Number of Snapshots $T$},
x label style = {at={(axis description cs:0.5,-0.1)},anchor=south},
ymode=log,
ymin=0.05,
ymax=50,
ylabel={\small RMSE (deg)},
y label style={at={(axis description cs:0.05,.5)},anchor=south},
yminorticks=true,
grid = both,
axis background/.style={fill=white},
legend columns=2, 
legend style={legend cell align=left, align=left, draw=white!15!black, row sep=-2pt, at={(0,0)}, anchor = south west, font = \scriptsize}
]
\addplot [color=mycolor12, line width = 1pt, mark = diamond, mark options={solid, mycolor12}, mark repeat = 2]
table[row sep=crcr]{%
10	32.7104655493504\\
14	32.6294792483341\\
21	32.0042963446547\\
30	31.3381252408657\\
43	30.6258839540993\\
62	28.8626605514792\\
89	27.0810165959826\\
127	24.2354884415446\\
183	19.1737729504103\\
264	13.9535949951435\\
379	8.24459596939816\\
546	4.05330116722897\\
785	1.56336541145223\\
1129	0.360435798318167\\
1624	0.186061234738674\\
2336	0.147641543298023\\
3360	0.1217579703069\\
4833	0.0984785643236317\\
6952	0.0825168455517376\\
10000	0.0673969986972146\\
};
\addlegendentry{MUSIC}

\addplot [color=mycolor11, line width = 1pt, mark = o, mark options={solid, mycolor11}, mark repeat = 2]
table[row sep=crcr]{%
10	24.3321660321569\\
14	24.3685007621054\\
21	24.9181747340657\\
30	24.882726905797\\
43	24.0170180464743\\
62	22.4223777067611\\
89	20.0551302649339\\
127	16.1469872280622\\
183	11.9502726650724\\
264	7.2743059444852\\
379	4.01270066482139\\
546	1.74021272011722\\
785	0.50394080705137\\
1129	0.200726754139633\\
1624	0.165952008289547\\
2336	0.136243435624157\\
3360	0.114419144129045\\
4833	0.0949254609539827\\
6952	0.0799551001976589\\
10000	0.0662213259141769\\
};
\addlegendentry{PR-DML}

\addplot [color=mycolor9, line width = 1pt, mark = square, mark options={solid, mycolor9}, mark repeat = 2]
table[row sep=crcr]{%
10	7.89518703340318\\
14	5.00020528319823\\
21	2.72975141734753\\
30	1.45600634978546\\
43	1.20486350079507\\
62	0.989469261213458\\
89	0.834542424486856\\
127	0.687961212145537\\
183	0.548349548149513\\
264	0.438376498393709\\
379	0.355022101223025\\
546	0.2939027214045\\
785	0.239123471881844\\
1129	0.19771354597633\\
1624	0.164659652500832\\
2336	0.135463911713817\\
3360	0.113867300976441\\
4833	0.0946103938939303\\
6952	0.0797149841333502\\
10000	0.0661080879564571\\
};
\addlegendentry{PR-DML-OLS}

\addplot [color=mycolor6, mark=asterisk, mark options={solid, mycolor6}, line width = 1pt, dashdotted, mark repeat = 2]
table[row sep=crcr]{%
10	9.0633688635083\\
14	7.12271627514963\\
21	4.42386471849575\\
30	2.6382680023076\\
43	1.47068633339881\\
62	1.0259980567186\\
89	0.797248020583462\\
127	0.644608652374695\\
183	0.508967097748769\\
264	0.418695236322513\\
379	0.34370206209646\\
546	0.285208313743732\\
785	0.234935896524986\\
1129	0.195573590994308\\
1624	0.162976134042405\\
2336	0.133941708166267\\
3360	0.112312609046248\\
4833	0.0931482785405665\\
6952	0.0780755196494338\\
10000	0.064328613152419\\
};
\addlegendentry{root-WSF}

\addplot [color=mycolor3, line width = 1pt, mark = triangle, mark options={solid, mycolor3}, mark repeat = 2, dashdotted]
table[row sep=crcr]{%
10	12.665433749657\\
14	10.3039668717057\\
21	8.33431276377532\\
30	7.13167602201765\\
43	6.6332019154563\\
62	6.42685895966661\\
89	6.23815905933418\\
127	6.17547164151683\\
183	6.11770535691691\\
264	6.05227574770552\\
379	6.01809688254871\\
546	5.9692315236454\\
785	5.91932585396126\\
1129	5.87553218638408\\
1624	5.84985866681513\\
2336	5.83046700588692\\
3360	5.82087925560228\\
4833	5.81714415061393\\
6952	5.81664410645765\\
10000	5.81693303179066\\
};
\addlegendentry{OMP}


\addplot [color=mycolor3, line width = 1pt,  dashdotted]
table[row sep=crcr]{%
	10	6.94966534069841\\
	14	4.29659231475543\\
	21	2.43302839077095\\
	30	1.77721900994899\\
	43	1.68673883463866\\
	62	1.69732462173736\\
	89	1.72592786781721\\
	127	1.74310609097307\\
	183	1.77523016269657\\
	264	1.79645476015829\\
	379	1.81527518680431\\
	546	1.83048679349454\\
	785	1.83778275706958\\
	1129	1.84374534128234\\
	1624	1.84938302070396\\
	2336	1.85074975302373\\
	3360	1.85320092036174\\
	4833	1.85405343395956\\
	6952	1.85400923802093\\
	10000	1.85476168897204\\
};
\addlegendentry{OLS}

\addplot [color=black, line width = 1pt]
table[row sep=crcr]{%
10	2.03842510471162\\
13	1.7878170848441\\
16	1.61151654263899\\
20	1.44138421448248\\
26	1.26417758421443\\
33	1.1221160891057\\
42	0.994649604194096\\
53	0.885435284438131\\
67	0.787512322340796\\
85	0.699174042754027\\
108	0.620273006464103\\
137	0.550724599496343\\
174	0.488675015267862\\
221	0.433609334123298\\
281	0.384540068555287\\
356	0.341640823758501\\
452	0.303197448281176\\
574	0.269053603570739\\
728	0.238907107203842\\
924	0.212060008137006\\
1172	0.188291603317438\\
1487	0.167162660686826\\
1887	0.148391426329669\\
2395	0.131717051514925\\
3039	0.116930931477341\\
3857	0.103793448128014\\
4894	0.0921430909969309\\
6210	0.0817991817273856\\
7880	0.0726158876659994\\
10000	0.0644606617055595\\
};
\addlegendentry{CRB}

\end{axis}

\end{tikzpicture}%
	\caption{Performance evaluation of the sequential DoA estimation techniques for four uncorrelated source signals at $\mb{\theta} = \left[90^\circ, 93^\circ, 135^\circ, 140^\circ\right]^\tT$ with an array composed of $M=10$ sensors and SNR = $3$dB. OMP and OLS are biased as the first DoA is estimated according to the conventional beamformer, which cannot resolve two closely-spaced sources at $90^\circ$ and $93^\circ$ regardless of the number of available snapshots $T$. On the other hand, PR-DML-OLS is asymptotically consistent and its RMSE is close to the CRB.}
	\label{figure:PerformanceSequential}
\end{figure}
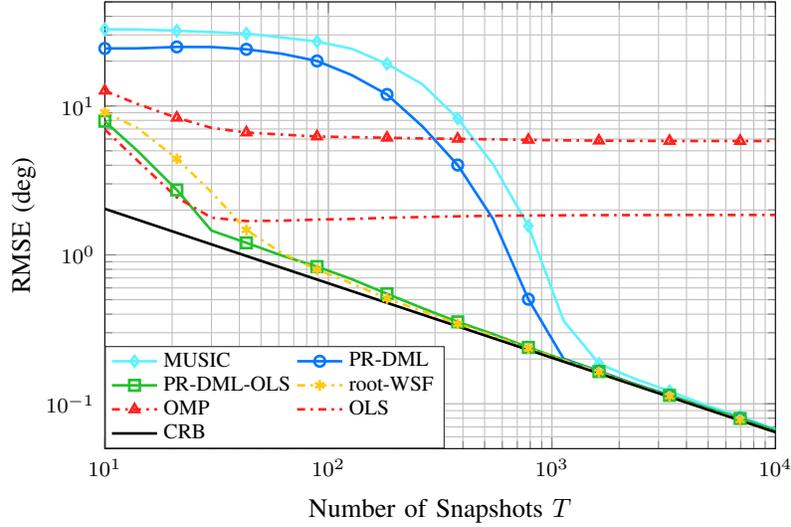

While the PR methods show excellent threshold performance in scenarios with low number of uncorrelated sources, their performance quickly degrades as the number of sources increases. This phenomenon can be explained in short as follows: the approximation error associated with the manifold relaxation of the interfering sources increases with the number of sources. The same holds true for the single source approximation methods. As the approximation error increases, the capability of incorporating the influence of multiple structured source signals in the optimization problem decreases, and thus a degradation in the estimation performance is observed. To overcome the degradation effect in scenarios with large source numbers, sequential estimation techniques have been proposed, in which the parameters of multiple sources are estimated one after the other. We revise three closely related and most widely known sequential estimation techniques: the Matching Pursuit (MP) technique, Orthogonal Matching Pursuit (OMP) and the Orthogonal Least-Squares (OLS)\cite{Chen89}, \cite{Ziskind_1988} method. 
These methods have in common, that the DoAs for $N$ sources are estimated sequentially and the approximation is successively improved. In each iteration, the DoA of one additional source is estimated based on minimizing a function approximation of a given multi-source criterion (c.f. Table~\ref{tab:Estimators}), while the source DoAs estimated in the previous iterations are kept fixed at the value of their respective estimates. Similar to the single source approximation, the remaining sources whose DoA estimates have not yet been determined are ignored in the optimization. The three methods differ, however, in the way the nuisance parameters corresponding to each source are treated in the optimization and in the corresponding parameter updating procedure. Concerning the MP algorithm, in each iteration, the nuisance parameters corresponding to the new source DoA are fixed and inserted as parameters in the objective in the following iterations. In contrast, in the OMP algorithm, the nuisance parameters of all estimated sources are updated in a refinement step after the DoA parameter of the current iteration is determined. The nuisance parameters are then inserted as parameters in the objective for the following estimation of the source DoA in the next iteration. The additional update step is generally associated with only a slight increase of the computational complexity. Nevertheless, this strategy effectively reduces error propagation effects. OLS yields further performance improvements at the cost of more sophisticated estimate and update expressions. More precisely, in the OLS algorithm, the nuisance parameters corresponding to the sources of the previous and current iterations are treated as variables and optimized along with the DoA parameter of the new source in the current iteration. In Table~\ref{tab:IterativeEstimator}, we provide the sequential estimation and update procedures of the MP, OMP, and OLS for the DML criterion (c.f. Table~\ref{tab:Estimators}). At this point, we remark that the sequential estimation approach is general and can also be applied to other multi-source criteria in Table~\ref{tab:Estimators}, hence the WSF and the COMET criteria. Furthermore, sequential estimation can also be combined with the concept of PR that we introduced in Section \ref{section:PR_methods} to further enhance the threshold performance and reduce error propagation effects. As an example, we also provide the PR-DML-OLS method in Table \ref{tab:IterativeEstimator}. A numerical performance comparison of the sequential estimators is provided in Figure~\ref{figure:PerformanceSequential}, where it can be observed that the OLS method shows improved performance as compared to OMP, however both methods suffer from a bias. The PR-DML-OLS method is, in contrast, asymptotically consistent and its Root Mean Square Error (RMSE) is close to the Cram\'{e}r-Rao bound (CRB).

\begin{table}
	\def\widthFirstCol{1.75}
	\def\widthSecondCol{7.55}
	\def\widthThirdCol{6.5}
	\def\widthSum{14.05}
	\def\widthMainLine{1}
	\def\widthSubLine{1}

	\begin{threeparttable}
		\caption{Sequential DoA Estimators}\label{tab:IterativeEstimator}
		\begin{tabular}{ccc}
			\noalign{\hrule height \widthMainLine pt}
			 & \small\textbf{Iterative DoA Estimation Step} & \small\textbf{Nuisance Parameters Update} \\ \noalign{\hrule height \widthMainLine pt}
			
			\multirow{2}{*}{\scriptsize \begin{tabular}[c]{@{}l@{}} \\[12pt]\textbf{MP}\end{tabular}} & \parbox{\widthSecondCol cm}{\scriptsize \begin{equation*}
					\iter{\hat{\theta}}{k} = \underset{\theta}{\arg\min}\text{ }\underset{\vec{s}\in\complexset^{T}}{\min} \norm{\vec{X} - \left[\hat{\vec{A}}^{(k-1)}, \vec{a}(\theta)\right]\begin{bmatrix}\hat{\vec{S}}^{(k-1)}\\ \vec{s}\trans\end{bmatrix}}_{\tF}^2
			\end{equation*}}  & \parbox{\widthThirdCol cm}{\scriptsize \begin{equation*}
				\iter{\hat{\vec{s}}}{k} = \underset{\vec{s}\in\complexset^{T}}{\arg\min} \norm{\vec{X} - \left[\hat{\vec{A}}^{(k-1)}, \vec{a}(\iter{\hat{\theta}}{k})\right]\begin{bmatrix}\hat{\vec{S}}^{(k-1)}\\ \vec{s}\trans\end{bmatrix}}_{\tF}^2
			\end{equation*}}\\
			& \parbox{\widthSecondCol cm}{\scriptsize \begin{equation*}
					\iter{\hat{\vec{A}}}{k} = \left[\iter{\hat{\vec{A}}}{k-1}, \vec{a}(\iter{\hat{\theta}}{k})\right]				
			\end{equation*}}  & \parbox{\widthThirdCol cm}{\scriptsize \begin{equation*}
					\iter{\hat{\vec{S}}}{k} = \begin{bmatrix}\hat{\vec{S}}^{(k-1)}\\ \hat{\vec{s}}^{(k)\tT}\end{bmatrix}
			\end{equation*}}\\
			\noalign{\hrule height \widthSubLine pt}
			
			\multirow{2}{*}{\scriptsize \begin{tabular}[c]{@{}l@{}} \\[6pt]\textbf{OMP}\end{tabular}} & \parbox{\widthSecondCol cm}{\scriptsize \begin{equation*}
					\iter{\hat{\theta}}{k} = \underset{\theta}{\arg\min}\text{ }\underset{\vec{s}\in\complexset^{T}}{\min} \norm{\vec{X} - \left[\hat{\vec{A}}^{(k-1)}, \vec{a}(\theta)\right]\begin{bmatrix}\hat{\vec{S}}^{(k-1)}\\ \vec{s}\trans\end{bmatrix}}_{\tF}^2
			\end{equation*}}  & \multirow{2}{*}{\scriptsize \begin{tabular}[c]{@{}l@{}} \parbox{\widthThirdCol cm}{\scriptsize \begin{equation*}
				\iter{\hat{\vec{S}}}{k} = \underset{\vec{S}\in\complexset^{k \times T}}{\arg\min} \norm{\vec{X} - \left[\hat{\vec{A}}^{(k-1)}, \vec{a}(\iter{\hat{\theta}}{k})\right]\vec{S}}_{\tF}^2
			\end{equation*}}\end{tabular}}\\
			& \parbox{\widthSecondCol cm}{\scriptsize \begin{equation*}
					\iter{\hat{\vec{A}}}{k} = \left[\iter{\hat{\vec{A}}}{k-1}, \vec{a}(\iter{\hat{\theta}}{k})\right]				
			\end{equation*}}  & \\
			\noalign{\hrule height \widthSubLine pt}
			
			\scriptsize\textbf{OLS} & \multicolumn{2}{c}{\parbox{\widthSum cm}{\scriptsize \begin{equation*}
				\iter{\hat{\theta}}{k} = \underset{\theta}{\arg\min}\text{ }\underset{\vec{S}\in\complexset^{k \times T}}{\min} \norm{\vec{X} - \left[\hat{\vec{A}}^{(k-1)}, \vec{a}(\theta)\right]\vec{S}}_{\tF}^2
			\end{equation*}
			\begin{equation*}
				\iter{\hat{\vec{A}}}{k} = \left[\iter{\hat{\vec{A}}}{k-1}, \vec{a}(\iter{\hat{\theta}}{k})\right]				
			\end{equation*}}} \\\noalign{\hrule height \widthSubLine pt}
			
			\scriptsize \textbf{PR-DML-OLS} & \multicolumn{2}{c}{\parbox{\widthSum cm}{\scriptsize \begin{equation*}
				\iter{\hat{\theta}}{k} = \underset{\theta}{\arg\min}\text{ }\underset{\substack{\vec{B}\in\complexset^{M\times (N-k)}\\ \vec{S}\in\complexset^{N\times T}}}{\min} \norm{\vec{X} - \left[\hat{\vec{A}}^{(k-1)}, \vec{a}(\theta), \vec{B}\right]\vec{S}}_{\tF}^2
			\end{equation*}
			\begin{equation*}
				\iter{\hat{\vec{A}}}{k} = \left[\iter{\hat{\vec{A}}}{k-1}, \vec{a}(\iter{\hat{\theta}}{k})\right]				
			\end{equation*}}} \\\noalign{\hrule height \widthSubLine pt}
		\end{tabular}
	\end{threeparttable}
	
\end{table}

\subsection{Sparse Reconstruction Methods}
The non-linear Least-Squares DML problem in Table \ref{tab:Estimators} generally requires a multidimensional grid-search over the parameter space to obtain the global minimum. More precisely, the objective function is evaluated at all possible combinations of $N$ DoAs on a particular discretized FoV. Clearly, the complexity of this brute-force multidimensional search strategy grows exponentially with the number of sources. To reduce the computational cost associated with the non-linear Least-Squares optimization, convex approximation methods based on sparse regularization have been proposed. 

\begin{figure}
	\centering
		\begin{tikzpicture}
	
	\scriptsize
	
	\begin{scope}[scale=0.4]		
		
		\tikzset{>=stealth'}
		
		\draw[lightgray,thin] (5,0) arc (0:180:5);
		\foreach \phi in {0,10,...,180} {
			\draw[lightgray,dashed,thin] (0,0) -- (\phi:5);
		}
		\draw[->,lightgray,thin] (173:5.7) arc (173:155:5.7);
		
		\foreach \x in {0,...,5} {
			\node[sensor, fill=black] (s\x) at (2*\x-5,0) {};
		}
		
		\draw[gray] (s0) -- (s1) -- (s2) -- (s3) -- (s4) -- (s5); 
		
		\node[circle, inner sep=2pt] (cs) at (0,0) {};
		
		\def\angA{40}
		\def\angB{100}
		\def\angC{130}
		
		\coordinate (src1) at (intersection cs: first line={(cs)--($(cs)+(\angA:6)$)}, second line={(0,5)--(3,5)}) {};
		\coordinate (src2) at (intersection cs: first line={(cs)--($(cs)+(\angB:6)$)}, second line={(0,5)--(3,5)}) {};
		\coordinate (src3) at (intersection cs: first line={(cs)--($(cs)+(\angC:6)$)}, second line={(0,5)--(3,5)}) {};
		
		\draw[->] (src1) -- (cs); 
		\draw[->] (src2) -- (cs); 
		\draw[->] (src3) -- (cs); 
		
		\node[source,fill=cs3,label={above:Source 3}] at (src1) {};
		\node[source,fill=cs2,label={above:Source 2}] at (src2) {};
		\node[source,fill=cs1,label={above:Source 1}] at (src3) {};
		
	\end{scope}
	
	\begin{scope}[xshift=4.9cm,yshift=0.5cm, scale = 1.3]
		
		\def\bSizeMod{1.5mm}
		\def\NSnp{4}
		
		\node[inner sep=0] (Y) at (0,0) {\drawMtx{\pgfmathdeclarerandomlist{color}{{gray}{lightgray}};}{6}{\NSnp}{\bSizeMod}};
		
		\node[inner sep=0, anchor=west] (eq) at ($(Y.east)+(2mm,0mm)$) {$=$};	
		\node[inner sep=0, anchor=west] (A) at ($(eq.east)+(2mm,0mm)$)
		{\drawMtx{\pgfmathdeclarerandomlist{color}{{red}{red}{orange}};}{6}{15}{\bSizeMod}};
		
		\node[inner sep=0, anchor=south west] (X) at ($(A.south east)+(2mm,0mm)$) {
			\begin{tikzpicture}
				
				\node[anchor=south west] at (0,11*\bSizeMod) {\drawMtxB{cs1}{1}{\NSnp}{\bSizeMod}};
				\node[anchor=south west] at (0,8*\bSizeMod) {\drawMtxB{cs2}{1}{\NSnp}{\bSizeMod}};
				\node[anchor=south west] at (0,3*\bSizeMod) {\drawMtxB{cs3}{1}{\NSnp}{\bSizeMod}};
				
				\draw (0,0) rectangle +(\NSnp*\bSizeMod, 15*\bSizeMod);
				
			\end{tikzpicture}	
		};	
		
		\node[anchor=north] (Y2) at ($(Y.south)+(0mm,-1mm)$) {$\mb{X}$};
		\node[anchor=center] (eq2) at (Y2.center-|eq.center) {$=$};
		\node[anchor=center] (A2) at (Y2.center-|A.center) {$\tilde{\mb{A}}$};
		\node[anchor=center] (X2) at (A2.center-|X.center) {$\sparse{\mb{S}}$};
		
	\end{scope}
	
\end{tikzpicture}
\caption{Concept of Sparse Reconstruction Methods. In the noiseless case, the received signal $\mb{X}$} is decomposable into a product of a fixed oversampled steering matrix $\tilde{\mb{A}}$ and a row-sparse source signal matrix $\tilde{\mb{S}}$. The locations of the non-zero rows in $\tilde{\mb{S}}$ correspond to the DoAs.
\label{figure:RowSparseReconstruction}
\end{figure}
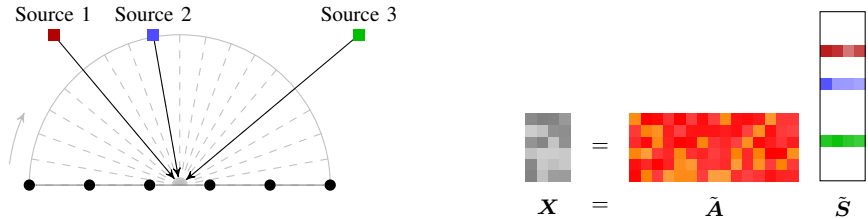

We assume that for a particular FoV discretization $\tilde{\mb{\theta}}\in\realset^K$ containing $K \gg M$ angles, a so-called oversampled dictionary matrix $\tilde{\mb{A}} =\mb{A}(\tilde{\mb{\theta}})$ of dimensions $M \times K$ is constructed and the DML problem is equivalently formulated as the linear least-squares minimization problem with a cardinality constraint, i.e.,
\begin{equation}\label{problem:L20constrained}
	\underset{ \tilde{\mb{S}} \in \mathbb{C}^{K \times T} }{\min}  \norm{ \mb{X} - \tilde{\mb{A}} \, \tilde{\mb{S}} }_{\tF}^2 \ \tst \ \quad \norm{ \tilde{\mb{S}} }_{2,0} \leq N,
\end{equation}
where $\| \mb{M} \|_{2,0} = \#\big(\{ k \ | \ \| \mb{m}_k \|_2 \neq 0 \} \big) $ denotes the $\ell_{2,0}$ mixed-pseudo-norm of a matrix $\mb{M} = [\mb{m}_1, \ldots, \mb{m}_K]^\tT$, i.e., the number of rows $\mb{m}_k$ with non-zero Euclidean norm $\|\mb{m}_k\|_2$, for $k = 1,\ldots, K$. This is illustrated in Figure~\ref{figure:RowSparseReconstruction}. Given a solution $\tilde{\mb{S}}^\star$ of the optimization problem in \eqref{problem:L20constrained}, the DoAs estimates $\hat{\mb{A}}(\hat{\mb{\theta}})$ are determined from the support of $\tilde{\mb{S}}^\star$, i.e. the locations of the nonzero rows.
Several approximation methods have been proposed to simplify the problem in \eqref{problem:L20constrained} using sparse regularization. Sparse regularization approaches are directly devised from the Lagrangian function of the optimization problem in \eqref{problem:L20constrained}, i.e,
\begin{equation}\label{problem:L20regularized}
\underset{\tilde{\mb{S}}}{\min}\norm{\mb{X} - \tilde{\mb{A}} \, \tilde{\mb{S}}}_{\tF}^2 + \mu \norm{\tilde{\mb{S}}}_{2,0},
\end{equation}
where the hyperparameter $\mu$ (also called regularization parameter) balances the trade-off between data matching and sparsity. For small values of $\mu$, the mismatch between the model and the measurements is emphasized in the minimization, whereas for larger value $\mu$, the row-sparsity of the solution is enhanced. Since the discretized FoV is given and thus the oversampled dictionary matrix $\tilde{\mb{A}}$ is constant, the data matching term in the objective function of \eqref{problem:L20regularized} is a simple linear LS function. Nevertheless, the sparse regularization term is both nonsmooth and nonconvex w.r.t. $\tilde{\mb{S}}$, and thus the problem in \eqref{problem:L20regularized} is difficult to solve directly. In \cite{Hyder10}, a convergent iterative fixed-point algorithm is proposed which solves a sequence of smooth approximation problems of \eqref{problem:L20regularized}.\\
In order to make the optimization in \eqref{problem:L20regularized} more tractable, a common approach is to convexify the regularizer in \eqref{problem:L20regularized} by approximating the $\ell_{2,0}$ norm by the closest convex mixed-norm $\ell_{2,1}$, which is defined as $\norm{\mb{M}}_{2, 1} = \sum_{k=1}^{K}\norm{\mb{m}_k}_2$ for $\mb{M} = \left[\mb{m}_1, \ldots, \mb{m}_K\right]\trans$. The resulting multiple measurement problem (MMP)
\begin{equation}\label{problem:MMP}
	\underset{\tilde{\mb{S}}}{\min} \norm{\mb{X} - \tilde{\mb{A}}\, \tilde{\mb{S}}}_\tF^2 + \mu \norm{\tilde{\mb{S}}}_{2, 1}
\end{equation}
is convex and thus can be solved efficiently \cite{Malioutov05}. One important drawback of the formulation in \eqref{problem:MMP} is that the number of optimization variables grows linearly with the number of snapshots $T$, and therefore also the associated computational complexity. Interestingly, the MMP can be equivalently expressed as the SPARse ROW-norm reconstruction (SPARROW) problem as follows \cite{Steffens18}			
	\begin{equation}\label{problem:SPARROW}
	\underset{{\tilde{\mb{D}}\in\mathcal{D}_{+}}}{\min}\text{tr}\left(\left(\tilde{\mb{A}}\, \tilde{\mb{D}} \, \tilde{\mb{A}}^\tH + \dfrac{\mu}{2\sqrt{T}} \vec{I}_M\right)^{-1}\hat{\vec{R}}\right) + \text{tr}\left(\tilde{\mb{D}}\right).
\end{equation}
We remark that in \eqref{problem:SPARROW}, the optimizing variable $\tilde{\mb{D}} =  \diag \big( \tilde{d}_1, \ldots, \tilde{d}_K \big)$ is a non-negative diagonal matrix, whose dimension does not depend on the number of snapshots $T$. The first term in \eqref{problem:SPARROW} can be interpreted as a data matching term while the second term induces sparsity.
The optima $\tilde{\mb{S}}^\star$ and $\tilde{\mb{D}}^\star$ of the respective problems share the same support and the diagonal elements of $\tilde{\mb{D}}^\star$ represent the scaled $\ell_2$ row-norms of $\tilde{\mb{S}}^\star$. The SPARROW problem is convex and can be efficiently solved using, e.g., a block-coordinate descent (BCD) algorithm. Remarkably, the support of the solution of \eqref{problem:SPARROW} and therefore also the solution of the MMP \eqref{problem:MMP} is fully encoded in the sample covariance matrix $\hat{\mb{R}}$ while the measurement matrix $\mb{X}$ is not explicitly required for the estimation of the DoAs. Making use of the Schur complement, the optimization problem in \eqref{problem:SPARROW} can further be reformulated as the semi-definite problem (SDP):
\begin{equation}\label{problem:SPARROW_SDP_oversample}
		\underset{\tilde{\mb{D}}, \vec{U}}{\min\text{ }}\tr\left(\vec{U}\hat{\mb{R}}\right) + \dfrac{1}{M}\tr\left(\tilde{\mb{D}} \right)\ 
		\tst \begin{bmatrix} \mb{U} & \mb{I}_M\herm\\
		\mb{I}_M & \tilde{\mb{A}} \tilde{\mb{D}} \tilde{\mb{A}}^\tH  + \dfrac{\mu}{2\sqrt{T}} \mb{I}_M
		\end{bmatrix}
		\succeq \mb{0}.
\end{equation}

An alternative SDP formulation also exists for the undersampled case, when the number of snapshots is smaller than the number of sensors, i.e., $M \geq T$. While from a computational point of view, the SDP formulations quickly become intractable when the dictionary $\tilde{\mb{A}}$ becomes large and the BCD solution is preferable for large $K$, the SDP formulations admit interesting extensions for uniform linear arrays (ULA) and other structured arrays that do not require the use of a sampling grid and the explicit formation of the dictionary $\tilde{\mb{A}}$. The so-called gridless sparse reconstruction methods are motivated by the following observation: in the ULA case, the dictionary $\tilde{\mb{A}}$ is a Vandermonde matrix so that the matrix product $\tilde{\mb{A}}\, \tilde{\mb{D}} \, \tilde{\mb{A}}^\tH$ with any diagonal matrix $\tilde{\mb{D}}$ can be substituted by a Toeplitz matrix $\text{Toep}\left(\mb{u}\right)$ with $\mb{u}$ denoting its first column. Inserting the compact Toeplitz reparameterization in the SDP problem 
\eqref{problem:SPARROW_SDP_oversample}
 and making use of the property $\tr\left( \tilde{\mb{D}} \right) = 1/M \tr\left( \tilde{\mb{A}}\, \tilde{\mb{D}} \, \tilde{\mb{A}}^\tH \right) = 1/M \tr\left( \text{Toep}\left(\mb{u}\right) \right)$, the SDP reformulation becomes independent of a particular choice of the dictionary $\tilde{\mb{A}}$. Consequently, the off-grid errors are avoided. An important question at this point is that, under which conditions the decomposition $\text{Toep}\left(\mb{u}\right) = \tilde{\mb{A}}\, \tilde{\mb{D}} \, \tilde{\mb{A}}^\tH$ holds, and whether the solution is unique? If such a decomposition exists with a unique solution, the gridless reformulation of the SDP is equivalent to the original, grid-based formulation. The answer to this question is provided by the well-known Carathéodory's theorem which states that, the Vandermonde decomposition of any positive semidefinite low rank Toeplitz matrix is always unique. Hence, provided that the solution $\text{Toep}\left(\mb{u}^\star \right)$ is positive semidefinite and rank deficient, it can be uniquely factorized as  $\text{Toep}\left(\mb{u}^\star \right) = \mb{A}^\star \mb{D}^\star (\mb{A}^\star)^\tH$ \cite{Yang_2016}. Given the generator vector $\mb{u}^\star$ retrieved from a low rank Toeplitz matrix, the DoA estimates can be uniquely recovered, e.g., by solving the corresponding system of linear equations.\\
We remark, that the gridless approach for sparse recovery in the MMP has first been introduced in the context of the atomic norm denoising problem \cite{Recht13}, which can be considered as the continuous angle equivalent of the $\ell_{2,1}$ norm regularized LS matching problem \eqref{problem:MMP}. The associated SDP formulation in the ULA case with Toeplitz parameterization can be shown to be equivalent to the gridless version of \eqref{problem:SPARROW_SDP_oversample}. \\
We further remark that gridless sparse reconstruction methods are not limited to contiguous ULA structures. Also other redundant array geometries can be exploited, such as shift invariant arrays or thinned ULAs, i.e., incomplete ULAs with missing sensors (``holes''). In thinned ULAs, ambiguities may arise in the array manifold and the model paramters may no longer be uniquely identifiable from the measurements. These ambiguities have, e.g., been characterized in \cite{Manikas1998}, \cite{Matter2022} and these references can provide guidelines for the choice of favorable thinned ULA geometries. Following a similar procedure as in the Toeplitz case, a substitution of the type $\mb{Q} = \tilde{\mb{A}}\, \tilde{\mb{D}} \, \tilde{\mb{A}}^\tH$ can be introduced where $\mb{Q}$ is no longer perfectly Toeplitz but contains other structured redundancies that can be expressed in the form of linear equality constraints in the problem \eqref{problem:SPARROW_SDP_oversample}. In these cases, ESPRIT or root-MUSIC can be employed to estimate the DoAs from the minimizer $\mb{Q}^\star$. Even though unique factorization guarantees for $\mb{Q}^\star$ similar to Carathéodory's theorem do not exist, the generalized gridless recovery approach performs well in practice as long as the number of redundant entries in $\mb{Q}$ is sufficiently large. \\
While we focused in our overview on sparse regularization methods that are based on the DML cost function in Table \ref{tab:Estimators} as the data matching term, there exists numerous alternative approaches that use other matching terms. See \cite{chellappa_chapter_2018} for an comprehensive overview on sparse DoA estimation techniques. A particularly interesting sparse DoA estimation methods is the SParse Iterative Covariance-based Estimation Approach (SPICE) \cite{Stoica11} which, as the name suggests, stems from a weighted version of the covariance matching criterion in Table \ref{tab:Estimators}. Remarkably, the SPICE formulation does not contain any hyperparameter to trade-off between the data matching quality versus the sparsity of the solution, which makes SPICE an attractive candidate among sparse reconstruction methods.

\begin{figure}
	\centering
%
\definecolor{mycolor1}{RGB}{136, 0, 21}%
\definecolor{mycolor2}{RGB}{185,122,187}%
\definecolor{mycolor3}{RGB}{255,32,27}%
\definecolor{mycolor4}{RGB}{255,174,201}%
\definecolor{mycolor5}{RGB}{255,137,29}%
\definecolor{mycolor6}{RGB}{255,201,14}%
\definecolor{mycolor7}{RGB}{255,242,0}%
\definecolor{mycolor8}{RGB}{239,228,176}%
\definecolor{mycolor9}{RGB}{31,194,43}%
\definecolor{mycolor10}{RGB}{181,230,29}%
\definecolor{mycolor11}{RGB}{0,119,251}%
\definecolor{mycolor12}{RGB}{94,243,255}%

\definecolor{mycolor13}{RGB}{63,72,204}%
\definecolor{mycolor14}{RGB}{112,146,190}%
\definecolor{mycolor15}{RGB}{163,73,164}%
\definecolor{mycolor16}{RGB}{200,191,231}%
\definecolor{mycolorGray}{RGB}{54, 69, 79}%

\begin{tikzpicture}
\begin{axis}[%
every tick label/.append style={font=\scriptsize},
xmode=log,
xmin=1,
xmax=1000,
xminorticks=true,
xlabel={\small Snapshots $T$},
xmajorgrids,
xminorgrids,
ymode=log,
ymin=0.05,
ymax=50,
yminorticks=true,
ylabel={\small RMSE},
ymajorgrids,
yminorgrids,
axis background/.style={fill=white},
legend columns = 2,
legend style={legend cell align=left, align=left, draw=white!15!black, row sep=-2pt, at={(0,0)}, anchor = south west, font = \scriptsize},
height=0.4\linewidth, 
width=0.6\linewidth, 
xlabel near ticks,
ylabel near ticks
]

\addplot [color=mycolor9, line width = 1pt, mark = square, mark options={solid, mycolor9}, solid]
  table[row sep=crcr]{%
1	25.2338099009377\\
2	16.6526858910026\\
3	11.7403720689031\\
5	5.02359331590936\\
7	2.5774970099818\\
10	1.14455492477441\\
12	1.01569852771116\\
15	0.899976150701936\\
20	0.799125156022095\\
30	0.624381284541233\\
50	0.486607916826029\\
70	0.420390421672202\\
100	0.34307226878249\\
200	0.262685865847341\\
300	0.216939011779134\\
500	0.180173017081253\\
1000	0.145629221396539\\
};
\addlegendentry{$\ell_{2,1}$-minimization in \eqref{problem:MMP}};

\addplot [color=mycolor11, line width = 1pt, mark = square, mark options={solid, mycolor11}, dashdotted, mark repeat = 2]
  table[row sep=crcr]{%
1	25.9176747495332\\
2	17.1294362291679\\
3	9.52364297091808\\
5	6.01983446018525\\
7	3.54877677862375\\
10	1.11265239592342\\
12	1.03517485420514\\
15	0.924172477713339\\
20	0.777378163203646\\
30	0.638161847728223\\
50	0.508135749296258\\
70	0.410378123714895\\
100	0.359965442846142\\
200	0.26565017856092\\
300	0.211758257649808\\
500	0.176473956609647\\
1000	0.138889585801956\\
};
\addlegendentry{Gridless-SPARROW};

\addplot [color=mycolor12!90!black, line width = 1pt, mark = diamond, mark options={solid, mycolor12!90!black}, mark repeat = 2]
  table[row sep=crcr]{%
1	31.9475873829783\\
2	27.5024827088506\\
3	27.647255346201\\
5	23.5478732278208\\
7	21.3904567131566\\
10	23.0374504455325\\
12	21.0692204753553\\
15	22.0703186215383\\
20	21.7752141979002\\
30	20.3770744393086\\
50	21.0247403467856\\
70	19.373569913917\\
100	18.165165942982\\
200	17.8770291818653\\
300	19.3059780395126\\
500	20.1356631383316\\
1000	19.9183439222989\\
};
\addlegendentry{MUSIC};

\addplot [color=TUDa-11b, line width = 1pt, mark = diamond, mark options={solid, TUDa-11b}, dashdotted, mark repeat = 2]
  table[row sep=crcr]{%
1	33.441380693537\\
2	30.0037226241956\\
3	28.4125008198382\\
5	26.0463309020916\\
7	23.4984485743261\\
10	26.0279634190038\\
12	25.2993372301518\\
15	25.6170038307142\\
20	23.9620319596515\\
30	24.4749187973479\\
50	23.2589909814172\\
70	21.8858213227379\\
100	22.3403044393838\\
200	22.9544173915294\\
300	22.8935553141523\\
500	23.0379905946774\\
1000	21.7947495625638\\
};
\addlegendentry{root-MUSIC};

\addplot [color=black,solid,line width = 1pt]
  table[row sep=crcr]{%
1	3.52061596670488\\
2	2.48945142401065\\
3	2.03262857609036\\
5	1.57446732484465\\
7	1.33066775852348\\
10	1.11331652215429\\
12	1.01631428804518\\
15	0.909019133829329\\
20	0.787233662422325\\
30	0.64277359375904\\
50	0.497890284802131\\
70	0.420794092588514\\
100	0.352061596670488\\
200	0.248945142401065\\
300	0.203262857609036\\
500	0.157446732484465\\
1000	0.111331652215429\\
};
\addlegendentry{CRB};

\end{axis}
\end{tikzpicture}%
\caption{Performance evaluation of the sparse reconstruction based DoA estimation techniques for two coherent sources at $\mb{\theta} = \left[90^\circ, 120^\circ\right]\trans$ with an array composed of $M = 6$ sensors and SNR = $10$ dB.}
\label{figure:PerformanceSparseRegularizationMethods}
\end{figure}
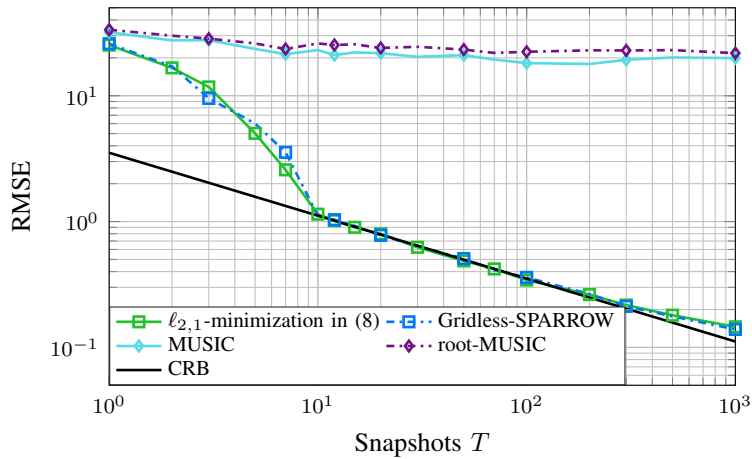

As mentioned above, the traditional super resolution methods, such as the multi-source estimation methods of Table \ref{tab:Estimators} as well as the PR methods and MUSIC for uncorrelated sources, are capable of resolving arbitrary closely spaced source even with a finite number of sensors as long as the number of snapshots or the SNR is sufficiently large. It is important to note that such guarantees generally do not exist in convex sparse optimization methods	
	\cite{Recht13}, \cite{Yang_2016}, \cite{chellappa_chapter_2018}.
Furthermore, sparse regularization-based DoA estimation methods are known to suffer from bias, which marks one of their most important drawbacks. First, the bias can be originated from the grid mismatch of the source DoAs in the formation of the dictionary. Second, the sparse regularization term generally introduces a bias to the solution. While the former source of bias can be entirely avoided in the gridless sparse reconstruction formulations, the latter remains and can only be reduced by decreasing the regularization parameter $\mu$, e.g., in \eqref{problem:L20regularized}. This in turn leads to an enlarged support set whose sizes is much larger the true number of sources $N$. However, in the context of sparse regularization-based DoA estimation, if the model order $N$ is known, it is often preferable to use comparably small values of the regularizers $\mu$ and to perform a local search for the $N$ largest maxima of the recovered row norm vector $\tilde{\mb{d}}^\star = \big[ \tilde{d}_1^\star, \ldots, \tilde{d}_K^\star \big]^\tT$ in \eqref{problem:SPARROW} to determine the DoA estimates. More specifically, 
the DoA estimates are the $N$ entries in the sample DoA vector $\tilde{\mb{\theta}}$ that are indexed by
$\big\{ i \big\}   = 	\underset{k}{^N\arg\max}\ \tilde{d}_k^\star$.\\
In conclusions, sparsity based methods have their merit in difficult scenarios with low sample size or highly correlated and even coherent source signals where the sample covariance matrix does not exhibit the full signal rank $N$. In these scenarios, conventional subspace based DoA estimation methods usually fail to resolve multiple sources. This is confirmed in the simulation example of Figure~\ref{figure:PerformanceSparseRegularizationMethods}. Furthermore, sparsity-based methods can resolve multiple sources even in the single snapshot case, provided that the scene is sparse in the sense that the number of sources is small compared to the number of sensors in the array. A simple but interesting theorem for robust sparse estimation with noisy measurements regardless of the chosen sparse estimation approach is given by \cite[Theorem 5]{Massoud10}.\\
Another benefit of sparse regularization methods is that, unlike parametric methods in DoA estimation, the knowledge of the number of sources $N$ is not required for the estimation of the DoAs. In turn, the number of sources is implicitly determined from the sparsity of the solution. However, sparse reconstruction methods, with the exception of the hyperparameter-free SPICE method \cite{Stoica11}, are usually sensitive to a proper choice of the sparse regularization parameter $\mu$. Furthermore, the associated computational complexity of sparse regularization methods, in particular for the SDP formulations in the grid-based and gridless cases, is higher than that of conventional subspace-based methods. \\
\section{Exploitation of incomplete structural information}
\begin{figure}
	\centering 
	\input{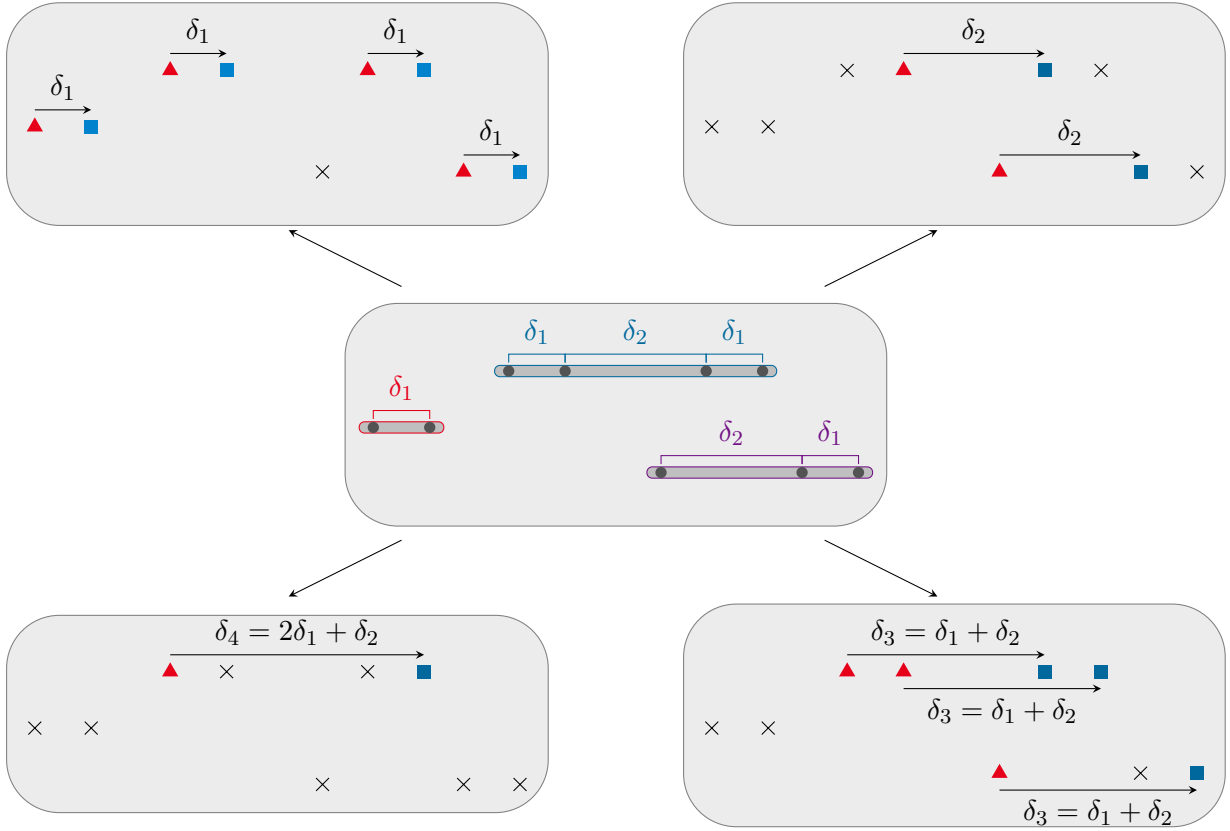}
	\caption{Equivalence between partly calibrated array setup and multiple shift-invariant setup. As depicted in the center, an examplary partly calibrated array setup comprises three linear subarrays with unknown intersubarray displacements. The displacements between sensors in one subarray are however a priori known. From the known intrasubarray displacement, multiple shift-invariant structures between sensor pairs are exploited while formulating the optimization problem. Such exploitation allows reinterpretation of the RARE algorithm \cite{Pesavento02,See04} as a generalized multiple shift ESPRIT \cite{Roy89}}
	\label{figure:PartlyCalibratedArray}
\end{figure}
In many modern applications such as networks of aerial base stations, DoA estimation is carried out in a distributed fashion from signals measured at multiple subarrays, where the exact locations of subarrays are often unknown. Even in conventional centralized large sensor arrays, it is a challenging task to have a central synchronized clock among all sensors of the device and to maintain precise phase synchronization due to large distances in the array. Therefore, in practice, large array systems are partitioned into local subarrays, where due to the proximity each subarray can be considered as perfectly calibrated, whereas the relative phase differences between subarrays are considered as unknown. In this setup, it is commonly assumed that the narrowband assumption remains valid, hence, the waveforms do not essentially decorrelate during the travel time over the array. 

DoA estimation in partly calibrated sensor arrays has first been considered in shift-invariant sensor array systems, which are comprised of two identically oriented identical subarrays separated by a known displacement $\delta$. For this configuration, the popular Estimation of Signal Parameters via Rotational Invariance Techniques (ESPRIT) algorithm has been proposed \cite{Roy89}. In shift-invariant arrays, the overall array steering matrix $\mb{A}(\mb{\theta})$ can be partitioned into two potentially overlapping blocks $\underline{\mb{A}}(\mb{\theta})\in\complexset^{M_1\times N}$ and $\overline{\mb{A}}(\mb{\theta})\in\complexset^{M_2\times N}$, respectively, representing the array response of the reference subarray and the shifted subarray. For notational simplicity, we assume that $M = 2M_1 = 2M_2$. Due to the shifting structure, the two subarray steering matrices are related though right-multiplication with a diagonal phase shift matrix $\mb{D}(\mb{z}^\delta) =
\diag \left(z_1^\delta, \cdots , z_N^\delta \right)$ with unit-magnitude generators $z_n = e^{-\tj\pi \cos(\theta_n)}$ that account for the known displacement $\delta$ measured in half-wavelength, hence $\overline{\mb{A}}(\mb{\theta}) = \underline{\mb{A}}(\mb{\theta}) \mb{D}(\mb{z})$.\\
Interestingly, ESPRIT as well as the enhanced Total-Least Squares ESPRIT method \cite{Roy89} can be reformulated as subspaces fitting techniques according to Table \ref{tab:Estimators}. Similar to the PR approach, a particular form of manifold relaxation is applied that maintains some part of the array structure and deliberately ignore other parts of the structure to admit a simple solution. The ESPRIT and TLS-ESPRIT estimators are obtained from minimizing the subspace fitting functions
 $  \big\| \hat{\mb{U}}_{\rm s} \mb{V}^{-1} - {\mb A}(\mb{\theta})\big\|_\tF^2 $ 
 \cite{Viberg91} and $ \big\| \hat{\mb{U}}_{\rm s} - {\mb A}(\mb{\theta})\mb{V}\big\|_\tF^2 $, respectively, for nonsingular $\mb{V}$, where the array manifold  $\mathcal{A}_N$
of the fully calibrated subarray defined in \eqref{eq:defManifold} is relaxed to the ESPRIT manifold ${\mathcal{A}}^{{\rm ESPRIT}}_{N}  = \big\lbrace\mb{A}\in \mathbb{C}^{M\times N}| \ \mb{A}(\mb{\vartheta}) = [\underline{\mb{A}}^\tT,\mb{D}(\mb{c})\underline{\mb{A}}^\tT   ]^\tT,\ \underline{\mb{A}}  \in \mathbb{C}^{M_1 \times N}, \mb{c} \in \mathbb{C}^N, \mb{\vartheta} = \cos^{-1} (-\frac{\arg(\mb{c})}{\pi\delta}) \big\rbrace$. We remark that in the ESPRIT manifold ${\mathcal{A}}^{{\rm ESPRIT}}_{N}$, only part of the shift-invariance structure of ${\mathcal{A}}_{N}$ is maintained and the particular subarray steering matrix structure in $\underline{\mb{A}}(\mb{\vartheta})=\left[\underline{\mb{a}}(\vartheta _1),\ldots,  \underline{\mb{a}}(\vartheta _N)\right]$ is relaxed to an arbitrary complex matrix $\underline{\mb{A}} \in \mathbb{C}^{M_1 \times N}$. In addition, the magnitude one structure of the diagonal shift matrix $\mb{D}(\mb{z}^\delta)$ is relaxed to an arbitrary diagonal matrix $\mb{D}(\mb{c})$. This implies that in the ESPRIT and TLS-ESPRIT algorithms, neither the subarray geometry nor potential directional gain factors between the two subarrays need to be known, as long as the subarrays are identical and subarray displacements are known. Due to this particular manifold relaxation, the subspace fitting problems admit efficient closed-form solutions.\\
The concept of DoA estimation in subarray structures has been generalized in \cite{Swindlehurst01} to cover the case of multiple shift invariance arrays.
In more general partly calibrated array scenarios, we assume that the sensor positions are generally unknown. Nevertheless, only several displacements between selected pairs of sensors in the array, the so-called \emph{lags} of the array are known. Let $\delta_1, \ldots, \delta_K$ denote known displacement in half-wavelength in the array that are all pointing into the same direction. This is illustrated in Figure~\ref{figure:PartlyCalibratedArray}. The subarrays are flexibly defined by pairs of sensors that share a common lag $\delta_k$ (or their summations). Depending on the number of known lags among the sensor arrays, one particular sensor can belong to one or more subarrays. For all known lags we consider again the subspace fitting approach and apply the ESPRIT manifold relaxation technique. Hence, defining $\mb{T} = \mb{V}^{-1}$ and relaxing the structure of the subarrays the objective becomes
$ \sum_{k = 1}^K \big\|  [  \hat{\underline{\mb{U}}}_{{\rm s},k} \mb{T}, \hat{\overline{\mb{U}}}_{{\rm s},k} \mb{T} ] - \underline{\mb A}_k\ [ \mb{I},\mb{D}({\mb z}^{\delta_k}) ] \big\|_\tF^2
$, where $\underline{\mb A}_k$ is an arbitrary complex-valued matrix of known dimension that models the unknown subarray structure corresponding to the $k$-th displacement $\delta_k$ and $\hat{\underline{\mb{U}}}_{{\rm s},k}$. The matrix $\hat{\overline{\mb{U}}}_{{\rm s},k}$ contains the corresponding rows of the signal eigenvectors in $\hat{\mb{U}}_{{\rm s},k}$. Inserting the LS minimizers $\hat{\underline{\mb{A}}}_{{\rm LS},k}  = \frac{1}{2} \left( \hat{\underline{\mb{U}}}_{{\rm s},k} \mb{T} +  \hat{\overline{\mb{U}}}_{{\rm s},k} \mb{T} \mb{D}^*(\mb{z}^{\delta_k}) \right)$ back into the objective function, the concentrated objective function of the relaxed multiple shift-invariant ESPRIT is given by
\begin{equation}
	\label{eq:concentratedMultShiftESPIRT}
 \sum_{k = 1}^K	\tr \big(-\mb{T}^\tH \hat{\underline{\mb{U}}}_{{\rm s},k}^\tH  \hat{\overline{\mb{U}}}_{{\rm s},k}  \mb{T} \mb{D}({\mb z}^{-\delta_k})+  \mb{T}^\tH (\hat{\underline{\mb{U}}}_{{\rm s},k}^\tH  \hat{\underline{\mb{U}}}_{{\rm s},k} +\hat{\overline{\mb{U}}}_{{\rm s},k}^\tH  \hat{\overline{\mb{U}}}_{{\rm s},k} )  \mb{T}   - \mb{T}^\tH \hat{\overline{\mb{U}}}_{{\rm s},k}^\tH  \hat{\underline{\mb{U}}}_{{\rm s},k}  \mb{T} \mb{D}({\mb z}^{\delta_k}\!)  \big).
\end{equation}
Due to the diagonal structure of $\mb{D}(\mb{z}^{\delta_k})$, the objective function in \eqref{eq:concentratedMultShiftESPIRT} is separable into $N$ identical terms, one for each source. Hence the subspace fitting problem reduces to finding the $N$ distinct minima of the RAnk REduction (RARE) \cite{Pesavento02,See04} function
$ f_{\rm RARE}(\theta) = 
	\min_{ \norm{\mb{t} } = 1}  
	\mb{t}^\tH \mb{M}(\theta)  \mb{t}$
with respect the DoAs $\theta \in \Theta$ where 
\begin{equation}
	\mb{M}(\theta)  = \sum_{k = 1}^{K}\left(  - \hat{\underline{\mb{U}}}_{{\rm s},k}^\tH  \hat{\overline{\mb{U}}}_{{\rm s},k} e^{\tj\pi\delta_k \cos(\theta_n)}
	+ \left(\hat{\underline{\mb{U}}}_{{\rm s},k}^\tH  \hat{\underline{\mb{U}}}_{{\rm s},k} +\hat{\overline{\mb{U}}}_{{\rm s},k}^\tH  \hat{\overline{\mb{U}}}_{{\rm s},k} \right)
	- \hat{\overline{\mb{U}}}_{{\rm s},k}^\tH  \hat{\underline{\mb{U}}}_{{\rm s},k}  e^{-\tj\pi\delta_k \cos(\theta_n)} 
	\right)
\end{equation} and
${\mb t}$ represents a particular column of $\mb{T}$. \footnote{The unit-norm constraint in the RARE problem is introduced to ensure that the zero solution $\mb{t}= \mb{0}$ is excluded, since the zero solution $\mb{t} = \mb{0}$ violates the constraint that $\mb{T}$ and $\mb{V}$ are nonsingular.} The minimization of the RARE cost function w.r.t. to the vector $\mb{t}$ admits the minor eigenvector of $\mb{M}(\theta)$ as a minimizer. As a result, the concentrated RARE cost function is given by $f_{\rm RARE}(\theta) = \lambda_{\min} \big(\mb{M}(\theta) \big)$. Thus, similar to the single-source approximation approach, the DoAs are determined from the $N$ deepest minima of the RARE function. This ensures that the corresponding transformation matrices $\mb{T}$ and $\mb{V}$ are nonsingular.
We remark that the RARE estimator has originally been derived from a relaxation of the MUSIC function and the minimum eigenvalue function can equivalently be replaced by the determinant of the matrix $\mb{M}(\theta)$. The latter is, e.g., useful for developing a search-free variant of the spectral RARE algorithm based on matrix polynomial rooting in the case that the shifts $\delta_1, \ldots, \delta_K$ are integer multiples of a common baseline. 

\section{Exploitation of array configuration}
As mentioned in Section~\ref{subsect:signal_model}, the number of signals $N$ that can be uniquely recovered from DoA estimation methods with second order statistics is strictly upper bounded by the Kruskal rank of the oversampled $(K \geq M)$ steering matrix $\tilde{\mb{A}} \in \mathcal{A}_K$, which e.g., for ULA geometries is equal to the number of sensors. A direct consequence is that, using the conventional signal model in Section~\ref{subsect:signal_model}, the number of uniquely identifiable sources $N$ must be less than the number of sensors $M$. If further information of the source signals is available, e.g., that the source signals are uncorrelated, then the number of uniquely identifiable source signals can be improved. 

This claim can be explained by comparing the number of equations and the number of unknowns which are implied from the covariance model $\mb{R} = \mb{A}(\mb{\theta}) \mb{D}\big(\mb{p}\big) \mb{A}^\tH(\mb{\theta}) + \nu \mb{I}_M$, with $\mb{D}\big(\mb{p}\big) = \diag ( p_1,\ldots, p_N )$. We assume that the number of snapshots $T$ is sufficiently high such that the covariance matrix $\mb{R}$ can be estimated with high accuracy. 
In addition, we remark that the structure of the covariance matrix depends on the geometry of the sensor array. For example, for a ULA, the covariance matrix is a Hermitan Toeplitz matrix. Conversely, if we assume that the sensor array does not exhibit any particular geometry, e.g., no ULA structure, then there is generally no relation between the elements in the covariance matrix. Consequently, the covariance matrix $\mb{R}$ is parameterized by maximally $M^2 - M + 1$ independent real-valued variables (note that the diagonal entries are all identical). Thus the number of independent equations from the covariance model is also $M^2 - M + 1$. On the other hand, in the uncorrelated source case, the model on the right hand side $\mb{A}(\mb{\theta}) \mb{D}\big(\mb{p}\big) \mb{A}^\tH(\mb{\theta}) + \nu \mb{I}_M$ contains only $2N + 1$ unknowns ($N$ DoA parameters in vector $\mb{\theta}$, $N$ source powers in $\mb{p} = [p_1, \ldots, p_N]^\tT$ and the noise variance $\nu$). This observation suggests that it is possible to significantly increase the number of uniquely identifiable sources in an array from $\mathcal{O}(M)$ to $\mathcal{O}(M^2)$ if the number of redundant entries in the covariance matrix is reduced. Therefore, from the viewpoint of improving the number of detectable sources for a fixed number of sensors, we should deviate from the conventional ULA array structure. The reason is that the covariance matrix in case of a ULA and uncorrelated source signals is a Hermitian Toeplitz matrix, which contains only $(2M-1)$ real-valued independent entries.\\
In fact, the covariance matching approach in Table \ref{tab:Estimators} combined with the concepts of sparse reconstruction in DoA estimation and positive definite Toeplitz matrix low-rank factorization has inspired an interesting line of research on nested and coprime arrays that aims at designing favorable non-redundant spatial sampling patterns \cite{Pal10}. These types of arrays include the class of minimum-redundancy and augmentable arrays whose design approaches rely on thinned uniform linear array geometries. One example is the sparse nonuniform arrays with inter-sensor spacings being integer multiples of a common baseline $\delta$. These geometries have the benefit over arrays with arbitrary non-integer spacings that they allow the use of search-free DoA estimation methods (c.f. the gridless sparse methods introduced in the previous section) and spatial smoothing techniques to build subspace estimates of the required rank. More precisely, given the stochastic signal model in the uncorrelated source case with source powers $\mb{p}$, hence, $ \mb{R} = \mb{A}(\mb{\theta}) \mb{D}\big(\mb{p}\big) \mb{A}^\tH(\mb{\theta}) + \nu \mb{I}_M$, an equivalent single snapshot model is obtained from vectorization. Defining $\mb{C}(\mb{\theta}) = \mb{A}^\ast(\mb{\theta}) \odot \mb{A}(\mb{\theta})$ as the steering matrix of a so-called virtual difference co-array, where  $\odot$ stands for the Khatri-Rao product, i.e., column-wise Kronecker product, the vectorized covariance model reads
$\mb{r} = {\rm vec}\big( \mb{ R} \big) = \mb{C}(\mb{\theta}) \mb{p} + \nu {\rm vec}\big(\mb{I}_M\big)$ \cite{Pal10}. For this model, the $\ell_{2,1}$-norm regularized LS approach in \eqref{problem:MMP} is a suitable candidate for DoA estimation. An interesting alternative approach that does not rely on sparsity but on the non-negativity property of the source power vector $\mb{p}$ is proposed in \cite{Qiao_2019}.
\\
In the vectorized covariance model, the number of identifiable sources is fundamentally limited by the Kruskal rank of the difference co-array steering matrix $\mb{C}(\mb{\theta})$. Hence, the design objective for the physical array is to place the sensors such that in the difference co-array, redundant rows of the difference co-array steering matrix $\mb{C}(\mb{\theta})$ are avoided and the number of contiguous lags is maximized. Avoiding redundant rows is equivalent to maximizing the diversity of the co-array, i.e., the number of different lags in the co-array. Maximizing the number of contiguous lags in the co-array corresponds to maximizing the size of the largest ``hole-free'' ULA partition of the co-array, which in turn is directly related to the Kruskal rank of the ULA partition (and therefore also the Kruskal rank of the entire difference co-array) due to the Vandermonde property of the ULA steering matrix. Because of the Khatri-Rao structure of the difference co-array steering matrix $\mb{C}(\mb{\theta})$, the Kruskal rank, and thus the number of uniquely identifiable sources, grows quadratically rather than linearly with the number of physical sensors $M$. While co-array designs allow to significantly increase the number of detectable sources for a given number of physical sensors using standard DoA estimation algorithms, recent theoretical performance results reveal that, in the regime where the number of sources $N$ exceeds the number of sensors $M$, the mean square estimation error of the MUSIC algorithm applied to the co-array data does not vanish asymptotically with SNR \cite{Wang_2017}.

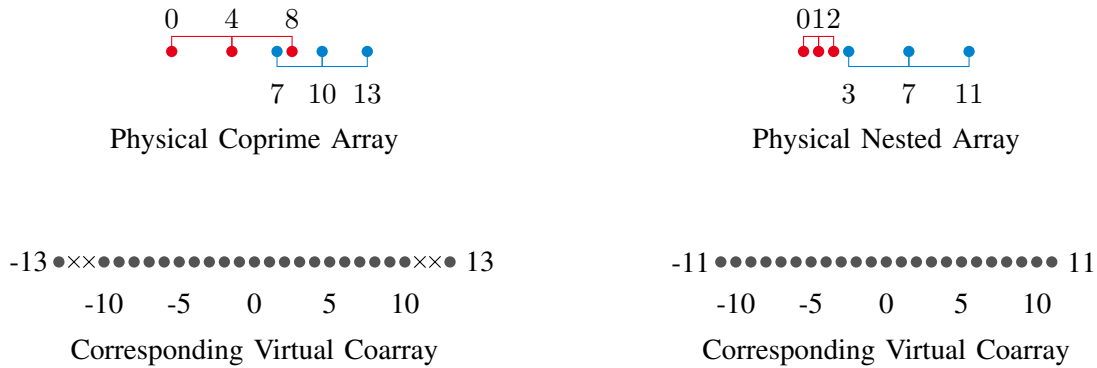
\begin{figure}
	\centering
	\begin{tikzpicture}
		
	\tikzset{>=stealth'}
	
	\begin{scope}[scale = 0.4]
	\def\initialSpacing{0.5};
	\coordinate (spacing1) at (2, 0);
	\coordinate (spacing2) at (1.5, 0);
	\coordinate (spacing3) at (2.75, -7);
	
	\coordinate (coords1) at (1.0, 0);
	\coordinate (coords2) at ($(coords1)+(spacing1)$);
	\coordinate (coords3) at ($(coords2)+(spacing1)$);
	
	\coordinate (coords4) at ($(coords1) + (spacing1) + (spacing2)$);
	\coordinate (coords5) at ($(coords4)+(spacing2)$);
	\coordinate (coords6) at ($(coords5)+(spacing2)$);
	
	\coordinate (coords7) at ($(coords1)+(spacing3)$);
	
	\coordinate (titleSpacing1) at (2.75, -3);
	\coordinate (coords8) at ($(coords1)+ (titleSpacing1)$);
	
	\coordinate (titleSpacing2) at (2.75, -10);
	\coordinate (coords9) at ($(coords1)+ (titleSpacing2)$);


	  \node[sensor, fill=TUDa-9b, label={[shift={( .0,.1)}]:$0$}] (s1) at (coords1) {};
	  \node[sensor, fill=TUDa-9b, label={[shift={(.0,.1)}]:$4$}] (s2) at (coords2) {};
	  \node[sensor, fill=TUDa-9b, label={[shift={( .0,.1)}]:$8$}] (s3) at (coords3) {};
	  
	  \node[sensor, fill=TUDa-2b, label={[shift={( .0,-0.9)}]:$7$}] (s4) at (coords4) {};
	  \node[sensor, fill=TUDa-2b, label={[shift={( .0,-0.9)}]:$10$}] (s5) at (coords5) {};
	  \node[sensor, fill=TUDa-2b, label={[shift={( .0,-0.9)}]:$13$}] (s6) at (coords6) {};
	  
  	  \node (title1) at (coords8) {Physical Coprime Array};

	  \def\r0{1.5mm};
	  \coordinate (sai1) at (0,0.5);
	  \coordinate (sai2) at (0,0.5);

	  \draw[TUDa-9b] ($(s1)+(0,\r0)$) -- ($(s1)+(sai1)$) -- node[midway, above]{} ($(s2) + (sai1)$)--($(s2)+(0,\r0)$);
	  \draw[TUDa-9b] ($(s2)+(0,\r0)$) -- ($(s2)+(sai1)$) -- node[midway, above]{} ($(s3) + (sai1)$)--($(s3)+(0,\r0)$);
	  \draw[TUDa-2b] ($(s4)-(0,\r0)$) -- ($(s4)-(sai2)$) -- node[midway, above]{} ($(s5) - (sai2)$)--($(s5)-(0,\r0)$);
	  \draw[TUDa-2b] ($(s5)-(0,\r0)$) -- ($(s5)-(sai2)$) -- node[midway, above]{} ($(s6) - (sai2)$)--($(s6)-(0,\r0)$);


    \foreach \iAnt in {-10,..., 10} {
    	\pgfmathparse{Mod(\iAnt,5)==0?1:0}
    	\ifnum\pgfmathresult>0
		  	\node[sensor, fill=TUDa-0d, label={[shift={( .0,-.9)}]:\iAnt}] (s1) at ($(coords7) + (\iAnt*\initialSpacing, 0)$) {};
		 \else
		  	\node[sensor, fill=TUDa-0d, label={[shift={( .0,-.9)}]:$$}] (s1) at ($(coords7) + (\iAnt*\initialSpacing, 0)$) {};
		 \fi
  	};
    \foreach \iAnt in {-12, -11, 11, 12} {
  	\node[cross, fill=TUDa-0d, label={[shift={( .0,-.9)}]:$$}, minimum size = 1.5mm] (s1) at ($(coords7) + (\iAnt*\initialSpacing, 0)$) {};
	  };
  \foreach \iAnt in {-13} {
  	\node[sensor, fill=TUDa-0d, label={[shift={(-0.4,-0.35)}]:\iAnt}] (s1) at ($(coords7) + (\iAnt*\initialSpacing, 0)$) {};
  };
\foreach \iAnt in {13} {
	\node[sensor, fill=TUDa-0d, label={[shift={(0.4,-0.35)}]:\iAnt}] (s1) at ($(coords7) + (\iAnt*\initialSpacing, 0)$) {};
};

\node (title2) at (coords9) {Corresponding Virtual Coarray};
	\end{scope}

\begin{scope}[scale = 0.4, xshift = 21cm, yshift = 0cm]
	
	\coordinate (spacing1) at (0.5, 0);
	\coordinate (spacing2) at (2, 0);
	\coordinate (spacing3) at (2.75, -7);
	\def\initialSpacing{0.5};
	
	\coordinate (coords1) at (1.0, 0);
	\coordinate (coords2) at ($(coords1)+(spacing1)$);
	\coordinate (coords3) at ($(coords2)+(spacing1)$);
	
	\coordinate (coords4) at ($(coords1)+(spacing2)-(spacing1)$);
	\coordinate (coords5) at ($(coords4)+(spacing2)$);
	\coordinate (coords6) at ($(coords5)+(spacing2)$);
	
	\coordinate (coords7) at ($(coords1)+(spacing3)$);
	\coordinate (titleSpacing1) at (2.75, -3);
	\coordinate (coords8) at ($(coords1)+ (titleSpacing1)$);
	
	\coordinate (titleSpacing2) at (2.75, -10);
	\coordinate (coords9) at ($(coords1)+ (titleSpacing2)$);
	
	
	\node[sensor, fill=TUDa-9b, label={[shift={( .0,.1)}]:$0$}] (s1) at (coords1) {};
	\node[sensor, fill=TUDa-9b, label={[shift={(.0,.1)}]:$1$}] (s2) at (coords2) {};
	\node[sensor, fill=TUDa-9b, label={[shift={( .0,.1)}]:$2$}] (s3) at (coords3) {};
	
	\node[sensor, fill=TUDa-2b, label={[shift={( .0,-0.9)}]:$3$}] (s4) at (coords4) {};
	\node[sensor, fill=TUDa-2b, label={[shift={( .0,-0.9)}]:$7$}] (s5) at (coords5) {};
	\node[sensor, fill=TUDa-2b, label={[shift={( .0,-0.9)}]:$11$}] (s6) at (coords6) {};
	
	\node (title1) at (coords8) {Physical Nested Array};

	\def\r0{1.5mm};
	\coordinate (sai1) at (0,0.5);
    \coordinate (sai2) at (0,0.5);

	\draw[TUDa-9b] ($(s1)+(0,\r0)$) -- ($(s1)+(sai1)$) -- node[midway, above]{} ($(s2) + (sai1)$)--($(s2)+(0,\r0)$);
	\draw[TUDa-9b] ($(s2)+(0,\r0)$) -- ($(s2)+(sai1)$) -- node[midway, above]{} ($(s3) + (sai1)$)--($(s3)+(0,\r0)$);
	
	\draw[TUDa-2b] ($(s4)-(0,\r0)$) -- ($(s4)-(sai2)$) -- node[midway, above]{} ($(s5) - (sai2)$)--($(s5)-(0,\r0)$);
	\draw[TUDa-2b] ($(s5)-(0,\r0)$) -- ($(s5)-(sai2)$) -- node[midway, above]{} ($(s6) - (sai2)$)--($(s6)-(0,\r0)$);


	\foreach \iAnt in {-10,..., 10} {
		\pgfmathparse{Mod(\iAnt,5)==0?1:0}
		\ifnum\pgfmathresult>0
		\node[sensor, fill=TUDa-0d, label={[shift={( .0,-.9)}]:\iAnt}] (s1) at ($(coords7) + (\iAnt*\initialSpacing, 0)$) {};
		\else
		\node[sensor, fill=TUDa-0d, label={[shift={( .0,-.9)}]:$$}] (s1) at ($(coords7) + (\iAnt*\initialSpacing, 0)$) {};
		\fi
	};
  \foreach \iAnt in {-11} {
	\node[sensor, fill=TUDa-0d, label={[shift={(-0.4,-0.35)}]:\iAnt}] (s1) at ($(coords7) + (\iAnt*\initialSpacing, 0)$) {};
};
\foreach \iAnt in {11} {
	\node[sensor, fill=TUDa-0d, label={[shift={(0.4,-0.35)}]:\iAnt}] (s1) at ($(coords7) + (\iAnt*\initialSpacing, 0)$) {};
};

\node (title2) at (coords9) {Corresponding Virtual Coarray};
\end{scope}
\end{tikzpicture}
	\caption{Examples of Coprime and Nested Array structure consisting of two subarrays, each with $3$ sensors. The baseline of each physical subarray is a multiple of the baseline of the virtual coarray.}	\label{figure:CoprimeArray}
\end{figure}
In the minimum redundancy array design, the number of contiguous lags in the difference co-array, hence the size of the largest ULA partion, is maximized by definition, which generally requires a computationally extensive combinatorial search over all possible spatial sampling patterns. The nested and coprime array designs, in contrast, represent systematic design approaches associated with computationally efficient analytic array design procedures \cite{Pal10}. 
Nested array and coprime arrays are composed of two uniform linear subarrays with different baselines. In the nested array structure each subarray is comprised of $M_1$ and $M_2$ sensors with baselines $\delta$ and $(M_1+1)\delta$, respectively, where the first sensor of the first subarray lies at the origin and the first sensor of the second subarray is displaced by $M_1$. 
It can be shown that with an equal split ($M_1 = M_2$ and $M_1 = M_2 +1$ for even and odd $M$, respectively) the difference co-array becomes a ULA with $2M_2(M_1+1)-1$ elements.\\
Coprime arrays represent more general array structures and comprise, as the name suggests, two uniform linear subarrays with $M_1$ and $M_2-1$ sensors, respectively, with $M_1$ and $M_2$ being coprime numbers. The first and the second subarray have baselines $L_1\delta$ and $L_2\delta$, respectively, where $L_1 = M_2$ and $L_2 = M_1/F$ are coprime numbers and integer $F$ is a given array compression factor in the range $1 \leq F \leq M_1$. Furthermore, the subarrays are displaced by integer multiples of the baseline $\delta$.\\
In Figure \ref{figure:CoprimeArray}, the nested and coprime array structures and their respective virtual co-arrays are illustrated for the case of $3$ sensors in each subarray. 
While the nested array structure yields a co-array with a maximum number of contiguous lags, the coprime array structure may often be preferable in practice as it can achieve not only a larger number of unique lags, i.e. degrees of freedom of up to $(M/2)^2+M/2$, but also a larger virtual co-array aperture, as well as a larger minimum inter-element spacing of the physical array to reduce, e.g., mutual coupling effects. 

In order to further increase the estimation performance of DoA estimators in a coprime array structure, low rank Toeplitz and Hankel matrix completion approaches have been proposed to fill the ``holes'' and augment the data in sparse virtual co-arrays to the corresponding full virtual ULA \cite{Chen_2014}. This concept has been successfully applied in \cite{Sun_2021} in the context of bistatic automotive radar to improve the angular resolution without increasing the hardware costs. Similarly, in \cite{Haghighatshoa_2017} matrix completion for data interpolation in coprime virtual arrays has been used for subspace estimation in hybrid analog and digital precoding with a reduced number of Analog-to-Digital converters and Radio Frequency chains in the hardware receives. Conditions under which in the noise-free case the completion from a single temporal snapshot is exact have been derived in \cite{Sarangi_2022}.


%

\section{Conclusions and future directions}
In this review paper, we revisit important developments in the area sensor array processing for DoA estimation in the past three decades from a modern optimization and structure exploitation perspective. From several illustrative examples, we show how novel concepts and algorithms that have advanced the research field in the last decades are proposed to solve, in some way or the other, the same notoriously challenging multi-source optimization problems, such as the well-known classical Deterministic Maximum Likelihood problem. Advances in convex optimization research and the development of efficient interior point solvers for semi-definite programs made it possible to compute close-to-optimal approximate solutions to these problems with significantly reduced effort. In addition, we also show how particular structure in the measurement model has been efficiently exploited to make the problems computationally tractable, both in terms of an affordable computational complexity, as well as in terms of well-posedness of the problem for identifying the parameters of interest. Nevertheless, we remark that our coverage of the sensor array processing research of the past three decades is by no means meant to be exhaustive. 
Given the long history of array signal processing, by now, this field of research can certainly be considered mature. Despite all the progress that has been made over the past decades, many important and fundamental research problems in this area have not yet been solved completely and require new ideas and concepts, and some of these are outlined below:

\emph{Harmonic retrieval in large dimensional data sets}: One example is the extension of the parameter estimation in one-dimensional (1D) spaces, such as in conventional DoA estimation, to higher dimensional spaces, e.g., as required in the aforementioned parametric MIMO channel estimation problem. With the trend to massive sensing systems and high dimensional data sets, the harmonic retrieval problem in extremely large dimensions gains significant interest. Due to the phenomenon known as curse-of-dimensionality where the computation workload increases exponentially with the number of dimensions, the extension of 1D DoA estimation methods to higher dimensions is not straightforward. Existing works on multidimensional harmonic retrieval either consider rather low dimensions or rely on dimensionality reduction approaches, i.e., projecting the multi-dimensional data sets into lower dimensions. This is, however, associated with a significant performance degradation if sources are not well separated in the projected domain.   

\emph{Incorporation of signal properties as prior information}: The use of particular structures in the array manifold, which is considered in this review article, is only one form of incorporating additional prior information into the estimation problem. As more information is exploited, the parameter estimation task can be correspondingly simplified and the estimation performance is enhanced. Theoretical investigation on the general use of additional side information incorporated in the DoA estimation problem as well as its estimation performance bound are addressed in \cite{Moore08}. In modern applications, the received signals as well as the waveforms often exhibit additional properties that can be exploited while designing novel DoA estimators. For example, constant modulus properties of the transmitted signals or signal waveforms with temporal dependence as, e.g., in radar chirp signals, enable coherent processing across multiple snapshots and dramatically enhance the resolution capabilities. Another example of signal exploitation is the DoA estimation with quantized or one-bit measurements, which has been studied in \cite{Sedighi21}. 

\emph{Robust sensor array processing}:  In many real-world applications, the classical array signal model may be over-simplistic. This can lead to a severe performance degradation of conventional high-resolution DoA estimation methods, which are known to be very sensitive to even small model mismatches. In recent years, significant efforts have been made to design DoA estimation methods that are robust to various model mismatch, including array imperfections due to miscalibration, impairments of the receiver frontends, mutual coupling between antennas, waveform decorrelation across the sensor array in inhomogeneous media and multipath environments, as well as impulsive and heavy-tailed noise \cite{Zoubir18}.

\emph{Combining model-based with data-driven DoA estimation}:
Recently, data-driven machine-learning approaches have been successfully introduced in many areas of signal processing to overcome existing limitations of traditional model-based approaches. Data-driven algorithms have the benefit that they naturally generalize to various statistics of the training data, and thus are flexible to adapt to time-varying estimation scenarios. As such, data-driven algorithms are potential candidates to overcome aforementioned challenges in DoA estimation. However, typical off-the-shelf data-driven algorithm are known to be data hungry, which limits their practical use in many DoA estimation applications. Recently, hybrid model-and-data-driven methods were proposed in the context of deep algorithm unfolding, which combine the benefits of both approaches. The hybrid algorithms inherit the structure of existing model-based algorithms in their learning architecture to reduce the number of learning parameters and therefore speed up the learning and improve the generalization capability of the algorithms \cite{Merkofer22}.
\section{Acknowledgement}
The work of Marius Pesavento was supported by the BMBF project Open6GHub under Grant 16KISK014 and the DFG PRIDE Project PE 2080/2-1 under Project No. 423747006

\addcontentsline{toc}{chapter}{Bibliography}
\bibliography{references}{}

\begin{thebibliography}{10}
\providecommand{\url}[1]{#1}
\csname url@samestyle\endcsname
\providecommand{\newblock}{\relax}
\providecommand{\bibinfo}[2]{#2}
\providecommand{\BIBentrySTDinterwordspacing}{\spaceskip=0pt\relax}
\providecommand{\BIBentryALTinterwordstretchfactor}{4}
\providecommand{\BIBentryALTinterwordspacing}{\spaceskip=\fontdimen2\font plus
\BIBentryALTinterwordstretchfactor\fontdimen3\font minus
  \fontdimen4\font\relax}
\providecommand{\BIBforeignlanguage}[2]{{%
\expandafter\ifx\csname l@#1\endcsname\relax
\typeout{** WARNING: IEEEtran.bst: No hyphenation pattern has been}%
\typeout{** loaded for the language `#1'. Using the pattern for}%
\typeout{** the default language instead.}%
\else
\language=\csname l@#1\endcsname
\fi
#2}}
\providecommand{\BIBdecl}{\relax}
\BIBdecl

\bibitem{Gao19}
F.~Gao, Z.~Tian, E.~G. Larsson, M.~Pesavento, and S.~Jin, ``{Introduction to
  the Special Issue on Array Signal Processing for Angular Models in Massive
  MIMO Communications},'' \emph{IEEE Journal of Selected Topics in Signal
  Processing}, vol.~13, no.~5, pp. 882--885, 2019.

\bibitem{VanTrees02}
H.~L. Van~Trees, \emph{{Optimum Array Processing}}.\hskip 1em plus 0.5em minus
  0.4em\relax Wiley, New York, 2002.

\bibitem{KrimV96}
H.~Krim and M.~Viberg, ``{Two Decades of Array Signal Processing Research: The
  Parametric Approach},'' \emph{IEEE Signal Processing Magazaine}, vol.~13,
  no.~4, pp. 67--94, 1996.

\bibitem{Trinh-Hoang18}
M.~Trinh-Hoang, M.~Viberg, and M.~Pesavento, ``{Partial Relaxation Approach: An
  Eigenvalue-Based DOA Estimator Framework},'' \emph{IEEE Transactions on
  Signal Processing}, vol.~66, no.~23, pp. 6190--6203, 2018.

\bibitem{Fuchs_1998}
J.-J. Fuchs, ``Detection and estimation of superimposed signals,'' in
  \emph{1998 {IEEE} {International} {Conference} on {Acoustics}, {Speech} and
  {Signal} {Processing} {(ICASSP)}}, vol.~3, May 1998, pp. 1649--1652.

\bibitem{Fuchs_2004}
------, ``On sparse representations in arbitrary redundant bases,'' \emph{IEEE
  Transactions on Information Theory}, vol.~50, no.~6, pp. 1341--1344, Jun.
  2004.

\bibitem{Ender10}
J.~H. Ender, ``{On Compressive Sensing Applied to Radar},'' \emph{Special
  Section on Statistical Signal and Array Processing, Signal Processing},
  vol.~90, no.~5, pp. 1402--1414, 2010.

\bibitem{Stoica11}
P.~Stoica, P.~Babu, and J.~Li, ``{SPICE: A Sparse Covariance-Based Estimation
  Method for Array Processing},'' \emph{IEEE Transactions on Signal
  Processing}, vol.~59, no.~2, pp. 629--638, 2011.

\bibitem{chellappa_chapter_2018}
Z.~Yang, J.~Li, P.~Stoica, and L.~Xie, ``{Chapter 11 - Sparse Methods for
  Direction-of-arrival Estimation},'' in \emph{Academic Press Library in Signal
  Processing, Volume 7}, R.~Chellappa and S.~Theodoridis, Eds.\hskip 1em plus
  0.5em minus 0.4em\relax Academic Press, 2018, pp. 509--581.

\bibitem{Malioutov05}
D.~Malioutov, M.~Cetin, and A.~Willsky, ``{A Sparse Signal Reconstruction
  Perspective for Source Localization with Sensor Arrays},'' \emph{IEEE
  Transactions on Signal Processing}, vol.~53, no.~8, pp. 3010--3022, 2005.

\bibitem{Steffens18}
C.~{Steffens}, M.~{Pesavento}, and M.~E. {Pfetsch}, ``{A Compact Formulation
  for the $\ell _{2,1}$ Mixed-Norm Minimization Problem},'' \emph{IEEE
  Transactions on Signal Processing}, vol.~66, no.~6, pp. 1483--1497, 2018.

\bibitem{Recht13}
G.~Tang, B.~N. Bhaskar, P.~Shah, and B.~Recht, ``{Compressed Sensing Off the
  Grid},'' \emph{IEEE Transactions on Information Theory}, vol.~59, no.~11, pp.
  7465--7490, 2013.

\bibitem{Zai16}
Z.~Yang, L.~Xie, and P.~Stoica, ``{Vandermonde Decomposition of Multilevel
  Toeplitz Matrices With Application to Multidimensional Super-Resolution},''
  \emph{IEEE Transactions on Information Theory}, vol.~62, no.~6, pp.
  3685--3701, 2016.

\bibitem{Scaglione08}
A.~Scaglione, R.~Pagliari, and H.~Krim, ``{The Decentralized Estimation of the
  Sample Covariance},'' in \emph{42nd Asilomar Conference on Signals, Systems
  and Computers}, 2008, pp. 1722--1726.

\bibitem{Pesavento02}
M.~Pesavento, A.~Gershman, and K.~Wong, ``{Direction Finding in Partly
  Calibrated Sensor Arrays Composed of Multiple Subarrays},'' \emph{IEEE
  Transactions on Signal Processing}, vol.~50, no.~9, pp. 2103--2115, 2002.

\bibitem{See04}
C.~See and A.~Gershman, ``{Direction-of-arrival Estimation in Partly Calibrated
  Subarray-based Sensor Arrays},'' \emph{IEEE Transactions on Signal
  Processing}, vol.~52, no.~2, pp. 329--338, 2004.

\bibitem{Moffet68}
A.~Moffet, ``{Minimum-redundancy Linear Arrays},'' \emph{IEEE Transactions on
  Antennas and Propagation}, vol.~16, no.~2, pp. 172--175, 1968.

\bibitem{Abramovich98}
Y.~Abramovich, D.~Gray, A.~Gorokhov, and N.~Spencer, ``{Positive-definite
  Toeplitz Completion in {DOA} Estimation for Nonuniform Linear Antenna Arrays.
  I. Fully Augmentable Arrays},'' \emph{IEEE Transactions on Signal
  Processing}, vol.~46, no.~9, pp. 2458--2471, 1998.

\bibitem{Pal10}
P.~Pal and P.~P. Vaidyanathan, ``{Nested Arrays: A Novel Approach to Array
  Processing With Enhanced Degrees of Freedom},'' \emph{IEEE Transactions on
  Signal Processing}, vol.~58, no.~8, pp. 4167--4181, 2010.

\bibitem{Tan14}
Z.~Tan, Y.~C. Eldar, and A.~Nehorai, ``{D}irection of {A}rrival {E}stimation
  {U}sing {C}o-{P}rime {A}rrays: {A} {S}uper {R}esolution {V}iewpoint,''
  \emph{IEEE Transactions on Signal Processing}, vol.~62, no.~21, pp.
  5565--5576, 2014.

\bibitem{Qin15}
S.~Qin, Y.~D. Zhang, and M.~G. Amin, ``{Generalized Coprime Array
  Configurations for Direction-of-Arrival Estimation},'' \emph{IEEE
  Transactions on Signal Processing}, vol.~63, no.~6, pp. 1377--1390, 2015.

\bibitem{Viberg91}
M.~Viberg and B.~Ottersten, ``{Sensor Array Processing based on Subspace
  Fitting},'' \emph{IEEE Transactions on Signal Processing}, vol.~39, no.~5,
  pp. 1110--1121, 1991.

\bibitem{Ottersten1998}
B.~Ottersten, P.~Stoica, and R.~Roy, ``{Covariance Matching Estimation
  Techniques for Array Signal Processing Applications},'' \emph{Digital Signal
  Processing}, vol.~8, no.~3, pp. 185--210, Jul. 1998.

\bibitem{Chen89}
S.~Chen, S.~A. Billings, and W.~Luo, ``{O}rthogonal {L}east {S}quares {M}ethods
  and {T}heir {A}pplication to {N}on-linear {S}ystem {I}dentification,''
  \emph{International Journal of Control}, vol.~50, no.~5, pp. 1873--1896,
  1989.

\bibitem{Ziskind_1988}
I.~Ziskind and M.~Wax, ``Maximum likelihood localization of multiple sources by
  alternating projection,'' \emph{IEEE Transactions on Acoustics, Speech, and
  Signal Processing}, vol.~36, no.~10, pp. 1553--1560, Oct. 1988.

\bibitem{Hyder10}
M.~M. Hyder and K.~Mahata, ``{Direction-of-Arrival Estimation Using a Mixed
  $\ell _{2,0}$ Norm Approximation},'' \emph{IEEE Transactions on Signal
  Processing}, vol.~58, no.~9, pp. 4646--4655, 2010.

\bibitem{Yang_2016}
Z.~Yang and L.~Xie, ``Exact {Joint} {Sparse} {Frequency} {Recovery} via
  {Optimization} {Methods},'' \emph{IEEE Transactions on Signal Processing},
  vol.~64, no.~19, pp. 5145--5157, Oct. 2016.

\bibitem{Manikas1998}
A.~Manikas and C.~Proukakis, ``Modeling and estimation of ambiguities in linear
  arrays,'' \emph{IEEE Transactions on Signal Processing}, vol.~46, no.~8, pp.
  2166--2179, 1998.

\bibitem{Matter2022}
F.~Matter, T.~Fischer, M.~Pesavento, and M.~E. Pfetsch, ``Ambiguities in doa
  estimation with linear arrays,'' \emph{IEEE Transactions on Signal
  Processing}, vol.~70, pp. 4395--4407, 2022.

\bibitem{Massoud10}
M.~Babaie-Zadeh and C.~Jutten, ``{On the Stable Recovery of the Sparsest
  Overcomplete Representations in Presence of Noise},'' \emph{IEEE Transactions
  on Signal Processing}, vol.~58, no.~10, pp. 5396--5400, 2010.

\bibitem{Roy89}
R.~Roy and T.~Kailath, ``{ESPRIT-Estimation of Signal Parameters via Rotational
  Invariance Techniques},'' \emph{IEEE Transactions on Acoustics, Speech, and
  Signal Processing}, vol.~37, no.~7, pp. 984--995, 1989.

\bibitem{Swindlehurst01}
A.~Swindlehurst, P.~Stoica, and M.~Jansson, ``{Exploiting Arrays with Multiple
  Invariances using MUSIC and MODE},'' \emph{IEEE Transactions on Signal
  Processing}, vol.~49, no.~11, pp. 2511--2521, 2001.

\bibitem{Qiao_2019}
H.~Qiao and P.~Pal, ``Guaranteed {Localization} of {More} {Sources} {Than}
  {Sensors} {With} {Finite} {Snapshots} in {Multiple} {Measurement} {Vector}
  {Models} {Using} {Difference} {Co}-{Arrays},'' \emph{IEEE Transactions on
  Signal Processing}, vol.~67, no.~22, pp. 5715--5729, Nov. 2019.

\bibitem{Wang_2017}
M.~Wang and A.~Nehorai, ``Coarrays, {MUSIC}, and the {Cramér}–{Rao}
  {Bound},'' \emph{IEEE Transactions on Signal Processing}, vol.~65, no.~4, pp.
  933--946, Feb. 2017.

\bibitem{Chen_2014}
Y.~Chen and Y.~Chi, ``Robust spectral compressed sensing via structured matrix
  completion,'' \emph{IEEE Transactions on Information Theory}, vol.~60,
  no.~10, pp. 6576--6601, 2014.

\bibitem{Sun_2021}
S.~Sun and Y.~D. Zhang, ``{4D} {Automotive} {Radar} {Sensing} for {Autonomous}
  {Vehicles}: {A} {Sparsity}-{Oriented} {Approach},'' \emph{IEEE Journal of
  Selected Topics in Signal Processing}, vol.~15, no.~4, pp. 879--891, Jun.
  2021.

\bibitem{Haghighatshoa_2017}
S.~Haghighatshoar and G.~Caire, ``Massive {MIMO} {Channel} {Subspace}
  {Estimation} {From} {Low}-{Dimensional} {Projections},'' \emph{IEEE
  Transactions on Signal Processing}, vol.~65, no.~2, pp. 303--318, Jan. 2017.

\bibitem{Sarangi_2022}
P.~Sarangi, M.~C. Hücümenoğlu, and P.~Pal, ``Single-{Snapshot} {Nested}
  {Virtual} {Array} {Completion}: {Necessary} and {Sufficient} {Conditions},''
  \emph{IEEE Signal Processing Letters}, vol.~29, pp. 2113--2117, Oct. 2022.

\bibitem{Moore08}
T.~J. Moore, B.~M. Sadler, and R.~J. Kozick, ``{Maximum-Likelihood Estimation,
  the Cram\'{e}r–Rao Bound, and the Method of Scoring With Parameter
  Constraints},'' \emph{IEEE Transactions on Signal Processing}, vol.~56,
  no.~3, pp. 895--908, Mar. 2008.

\bibitem{Sedighi21}
S.~Sedighi, B.~S. Mysore~R, M.~Soltanalian, and B.~Ottersten, ``{On the
  Performance of One-Bit DoA Estimation via Sparse Linear Arrays},'' \emph{IEEE
  Transactions on Signal Processing}, vol.~69, pp. 6165--6182, 2021.

\bibitem{Zoubir18}
A.~M. Zoubir, V.~Koivunen, E.~Ollila, and M.~Muma, \emph{Robustness in Sensor
  Array Processing}.\hskip 1em plus 0.5em minus 0.4em\relax Cambridge
  University Press, 2018, p. 125–146.

\bibitem{Merkofer22}
J.~P. Merkofer, G.~Revach, N.~Shlezinger, and R.~J.~G. van Sloun, ``{Deep
  Augmented Music Algorithm for Data-Driven DOA Estimation},'' in \emph{2022
  IEEE International Conference on Acoustics, Speech and Signal Processing
  (ICASSP)}, 2022, pp. 3598--3602.

\end{thebibliography}
\bibliographystyle{ieeetran}

\newpage

%
%
%
%

\vfill

\end{document}